\documentclass[numberedappendix]{emulateapj}

\begin{document}

\title{The DEEP2 Galaxy Redshift Survey: The Galaxy Luminosity Function to $z
\sim$ 1\altaffilmark{1}}  
\author{
  C.N.A. Willmer\altaffilmark{2,3},
  S. M. Faber \altaffilmark{2},
  D. C. Koo\altaffilmark{2},  
  B. J. Weiner\altaffilmark{2, 4}, 
  J. A. Newman\altaffilmark{5, 6},
  A. L. Coil,\altaffilmark{7}
  A. J. Connolly\altaffilmark{8},
  C. Conroy\altaffilmark{7},
  M. C. Cooper\altaffilmark{7},
  M. Davis\altaffilmark{7, 9},
  D. P. Finkbeiner\altaffilmark{10},
  B. F. Gerke\altaffilmark{9},
  P. Guhathakurta\altaffilmark{2}, 
  J. Harker\altaffilmark{2},
  N. Kaiser\altaffilmark{11},
  S. Kassin\altaffilmark{2}, 
  N. P. Konidaris\altaffilmark{2},
  L. Lin\altaffilmark{2,12},
  G. Luppino\altaffilmark{11},
  D. S. Madgwick\altaffilmark{5, 6},
  K. G. Noeske\altaffilmark{2},
  A. C. Phillips\altaffilmark{2}, 
  R. Yan\altaffilmark{7}.
}
\altaffiltext{1}{Based on observations taken at the W. M. Keck
  Observatory which is operated jointly by the University of
  California and the California Institute of Technology}
\altaffiltext{2}{UCO/Lick Observatory, University of California, 1156
  High Street, Santa Cruz, CA, 95064, {\tt cnaw@ucolick.org, faber@ucolick.org,
  koo@ucolick.org, bjw@ucolick.org, raja@ucolick.org,
  jharker@ucolick.org, kassin@ucolick.org, npk@ucolick.org,
  lihwai@ucolick.org, kai@ucolick.org, phillips@ucolick.org}}
\altaffiltext{3}{On leave from Observat\'orio Nacional, Rio de
  Janeiro, Brazil}
\altaffiltext{4}{Present address: Department of Astronomy, University
of Maryland, College Park, MD 20742}
\altaffiltext{5}{Lawrence Berkeley Laboratory, Berkeley, CA 94720, {\tt
    janewman@lbl.gov}}
\altaffiltext{6}{Hubble Fellow}
\altaffiltext{7}{Department of Astronomy, University of California, 601
  Campbell, Berkeley, CA 94720, {\tt acoil@astron.berkeley.edu,
  cconroy@astron.berkeley.edu, cooper@astron.berkeley.edu,
  mdavis@astron.berkeley.edu,
  renbin@astron.berkeley.edu}}
\altaffiltext{8}{Department of Physics and Astronomy, University of
  Pittsburgh, Pittsburgh, PA 15260, {\tt ajc@phyast.pitt.edu}}
\altaffiltext{9} {Department of Physics, Le Conte Hall, 
UC-Berkeley, Berkeley, CA 94720, bgerke@astro.Berkeley.EDU}
\altaffiltext{10} {Department of Astrophysics, Princeton University,
  Peyton Hall, Princeton, NJ 08544 {\tt dfink@astro.princeton.edu }}
 \altaffiltext{11} {Institute for Astronomy, 2680 Woodlawn Drive 
Honolulu, HI, 96822-1897, {\tt kaiser@ifa.hawaii.edu, ger@ifa.hawaii.edu}}
\altaffiltext{12} {Department of Physics, National Taiwan University,
  No. 1, Sec.4, Roosevelt Road, Taipei 106, Taiwan}


\begin{abstract}
The evolution of the $B$-band galaxy luminosity function is measured using a
sample of more than 11,000 galaxies with spectroscopic redshifts from
the DEEP2 Redshift Survey. The rest-frame $M_B$ versus  $U-B$
color-magnitude diagram of DEEP2 galaxies shows that the
color-magnitude bimodality seen in galaxies locally is still present
at redshifts $z >$ 1.
Dividing the sample at the trough of this color bimodality into
predominantly red and blue galaxies, we find that the luminosity
function of each galaxy color type evolves differently.
Blue counts tend to shift to brighter magnitudes at constant number
density, while the red counts remain 
largely constant at a fixed absolute magnitude.
Using Schechter functions with fixed faint-end slopes
we find that $M^*_B$ for blue galaxies brightens by $\sim$ 1.3 $\pm$0.14
magnitudes per unit redshift, with no significant evolution in number density.
For red galaxies $M^*_B$ brightens somewhat less
with redshift, while the formal value of $\phi^*$ declines.
When the population of blue galaxies is subdivided into two halves
using the rest-frame color as the criterion, the measured evolution of
both blue subpopulations is very similar.

\end{abstract}

\clearpage

\section{Introduction}

The luminosity function is an important tool to analyze redshift
surveys since it provides a direct estimate of how much light is
contained in galaxies. 
By characterizing the observed changes with redshift in the luminosity
function of galaxies as a function of (rest-frame) wavelength, it is
possible to measure how the star-formation rates (e.g., using
ultra-violet data) and stellar masses (e.g., using $K$-band data)
have changed as a function of time. These analyses quantify the
observed changes undergone by the galaxies' masses and mass-to-light
ratios, thus providing valuable data for theories of galaxy formation.

The measurement of the galaxy luminosity function for samples of field
galaxies (i.e., galaxies selected for redshift measurements
independent of their local environment) has been made for almost every major
redshift survey (see Binggeli et al. 1998, Tresse 1999 and de
Lapparent et al. 2003 for reviews). 
Thanks to major
surveys such as the Two-Degree Field Galaxy Redshift Survey (Colless
et al. 2001) and the Sloan Digital Sky Survey (York et al. 2000),
precise measurements of the luminosity function in the local ( $z <
0.3$) universe are available (e.g., Norberg et al. 2002; Blanton et
al. 2003; Bell et al. 2003), providing a benchmark to measure the
luminosity function evolution.

The characterization of properties of galaxies at redshifts $z \sim$
1, a time when the universe was half its present age, is then
an important step to fully understand how galaxies formed and evolve.
The DEEP2 Redshift Survey (Davis et al. 2003) is a project that is
measuring 50,000 galaxy redshifts in four widely separated regions of
the sky, comprising a total area of 3.5 $\sq\degr$, and is specifically
designed to probe the properties of galaxies at redshifts beyond $z$ =
0.7. 

In a series of two papers, the $B$-band galaxy luminosity function to $z\sim$1
will be investigated  using the DEEP2 Redshift Survey (this
paper), followed by an analysis that combines DEEP2 with other
major current surveys of distant galaxies
(Wolf et al. 2003; Gabasch et al. 2004; Ilbert et al. 2005),
 discussing possible
evolutionary scenarios for early and late type galaxies (Faber et
al. 2005, hereafter Paper II). The choice of  $B$-band
rather than other rest-frame wavelengths is motivated by the large
number of measurements in this spectral range both for local as well
as for distant samples of galaxies. 
An additional advantage is that
for most of the higher-redshift intervals considered in this work,  
observed $R$ and $I$ are sampling rest-frame $B$, thus minimizing the
importance of K-corrections.
The present paper uses data from $\sim$ 1/4 of the total DEEP2 survey
 to
measure the galaxy luminosity function, and discusses the importance of
several selection effects in its measurement. 
This analysis will also take advantage of the recently found 
$bimodality$ of galaxies in the color-magnitude diagram 
(Strateva et al. 2001; Hogg et al. 2003; Baldry et al. 2004
and references therein), where the predominantly red early-type
galaxies occupy a distinct locus in color from
the blue star-forming galaxies. This bimodality has been shown to
extend to $z\sim$1 (Im et al. 2002; Bell et al. 2004; Weiner et
al. 2005) and beyond (e.g., Giallongo et al. 2005).  A bimodal
distribution is also seen for other parameters 
such as spectral class (Madgwick et al. 2002, 2003), morphologies,
metallicities, and star formation rates (Kauffmann et al. 2003).
The present paper will show that this bimodality persists to
$z \sim$ 1, and that it is related on how these two populations have evolved
over the last 6 Gyr. 

%
This paper is organized as follows: \S2 presents the DEEP2 data used
in this paper, describing the selection effects that are present
in this sample; \S3 describes the methods used to measure the
luminosity function and its evolution, and the weighting scheme that
was adopted to correct for the incomplete data sampling; \S4
presents the analysis of DEEP2 data showing how the evolution of the
galaxy luminosity function depends on the internal properties of
galaxies, blue galaxies showing mainly luminosity evolution 
while the red galaxy luminosity function shows a decrease in number
density toward higher redshifts.
Two appendices follow, describing the method used
to calculate the K-corrections and another estimating biases in the
luminosity function calculation by making cuts at different limiting
absolute magnitudes.
Throughout this work, a ($H_0, \Omega_M, \Omega_\Lambda)$ = (70, 0.3, 0.7)
cosmology is used. Unless indicated otherwise, magnitudes and
colors are converted into the Vega system, following the relations
shown in Table 1. 

%
%
\section{Data}
This section gives a brief description of the DEEP2 data;
for more details the reader is  referred to
Davis et al. (2003), who give an outline of the project, Faber et
al. (in preparation), who describe the survey strategy and
spectroscopic observations, Coil et al. (2004b), who describe the
preparation of the source catalog, and Newman et al. (in preparation),
where the spectroscopic reduction pipeline is described. 

The photometric catalog for DEEP2 (Coil et al. 2004b)
is derived from Canada-France-Hawaii 
Telescope (CFHT) images taken with the 12K $\times$ 8K mosaic camera
(Cuillandre et al. 2001)  in $B$, $R$ and $I$ in four different regions
of the sky. 
The $R$-band images have the highest signal-to-noise and were 
used to define the galaxy sample, which has a limiting
magnitude for image detection at $R_{AB}\sim 25.5$. 
Objects were identified using the $imcat$ software written
by N. Kaiser and described by Kaiser, Squires \& Broadhurst (1995).
In addition to magnitudes, $imcat$ calculates other image parameters
which are used in the object classification.
The separation between stars and galaxies is based on
magnitudes, sizes, and colors, which are used to
assign each object a probability of being a galaxy ($P_{gal}$). For the
DEEP2 fields, the cut is made at $P_{gal} >$ 0.2, $i.e.$, objects are
considered as part of the sample whenever the probability of being a
galaxy is greater than 20\%.
In Fields 2, 3, and 4, the spectroscopic sample
is pre-selected using $B$, $R$, and $I$ to have
estimated redshifts greater than 0.7, which 
approximately doubles the efficiency of the survey
for galaxies near $ z \sim 1$.  The fourth field,
the Extended Groth Strip (EGS), does not have
this pre-selection but instead has roughly equal
numbers of galaxies below and above $ z = 0.7$ 
selected using a well understood algorithm.
In addition to the redshift pre-selection, 
a surface brightness cut defined as
\begin{equation}
SB = R_{AB} + 2.5 Log_{10}\lbrace\pi (3r_g)^2 \rbrace \leq 26.5,
\end{equation}
is applied when selecting spectroscopic candidates, where $R_{AB}$ is
the $R$-band (AB) magnitude, and $r_g$ is the 1 $\sigma$ radius of the
Gaussian fit to the image profile in the CFHT photometry;
the minimum size for 
$r_g$ is fixed at 0$''$.33, so that for compact objects with $3r_g < 1''$, the surface brightness is measured within a circular aperture of
1$''$.  Finally, galaxies were selected  to lie within
bright and faint apparent-magnitude cuts of 18.5$~\leq~R_{AB}~\leq$~24.1.

The DEEP2 sample used here combines data from the
first season of observations in Fields 2, 3, and 4
with about 1/4 of the total EGS data, which
provides an initial
sample at low redshifts. The total number of
galaxies is 11284, with 4946 (45\%) in EGS,
3948 (36\%) in Field 4, 2299 (21\%)
in Field 3, and 91 (1\%) in Field 2.  
Because of the $BRI$ redshift pre-selection, for $z < 0.8$,
only EGS is sampled well enough to be used,
while data in all four fields are used for $z \ge 0.8$.
   
DEEP2 spectra were acquired with 
the DEIMOS spectrograph (Faber et al. 2003) on the Keck 2
telescope and processed by an automated pipeline that does the
standard image reduction (division by flatfield, rectification of
spectra, extraction of 2-D and 1-D spectra) and redshift
determination (Newman et al., in preparation). 
The only human intervention occurs during redshift validation,
where spectra and redshifts are visually examined, and redshifts are
given a quality assessment that ranges from 1 (for completely indeterminate) to
4 (for ironclad).  Only redshifts with quality 3 and 4
are used in this paper, which means that two or more
features have been identified (the [O II] $\lambda$3727
doublet counts as two features).  Duplicate
observations and other tests indicate
an rms accuracy of 30 km s$^{-1}$ and an unrecoverable  failure
rate of $\sim$1\% for this sample.  

The apparent color-magnitude (CM) diagram in $R$ versus $R-I$ 
is shown in Figure 1$a$
for the DEEP2 parent catalog after applying the
photometric-redshift cut in three of the fields and converting into
Vega magnitudes (cf. Table 1). Figure 1$b$ 
shows the distribution of galaxies placed on masks, Figure 1$c$
shows galaxies with successful redshifts, 
and Figure 1$d$ shows galaxies with ``failed'' redshifts.
Although failures are found in all parts of the diagram,
the largest concentration is at faint and blue magnitudes.
Independent data show that the great majority
of these are beyond $z~\sim~1.4$
(C. Steidel, private communication), corresponding to
[O II] $\lambda$3727 passing beyond the DEEP2
wavelength window at that redshift.
Redshift histograms corresponding to the
rectangular regions outlined in Figure 1
are shown in Figure 2, where the vertical
bars at the right of each diagram represent the number of failed
redshifts in each bin. The increase in failures for faint and blue
galaxies is apparent. 

Figure 3$a$ plots $U-B$ versus distance for
the whole sample, where the rest-frame color is calculated using
the K-correction procedure described in Appendix A. Throughout this
paper, the rest-frame  colors and magnitudes are corrected for Galactic
extinction (Schlegel, Finkbeiner \& Davis 1998) but not internal extinction.
Color bimodality dividing red and blue galaxies
is immediately apparent, extending to beyond $z = 1$.
Panel $b$ shows the EGS by itself, 
while panels $c$ and $d$ show the high effectiveness of the
$BRI$ photometric selection  in Fields 2, 3, and 4. 

Figure 4 plots CM diagrams using $U-B$ versus $M_B$ as a function of redshift. 
The solid line in each panel represents the limiting absolute magnitude 
at the high redshift end of each bin. 
The slope of this line changes with redshift 
because of the adoption of a fixed apparent magnitude
limit ($R$) for the sample, with the color-redshift-dependence of the
K-correction.  At  $z \sim$ 0.4 the $R$-band filter used to select
the sample coincides with rest $B$ but
differs from it increasingly as the redshift is
either greater or smaller than 0.4. 

The bimodality in color-magnitude distribution is clearly seen; while
red galaxies tend to be brighter on average than blue galaxies, it is
clearly seen that blue galaxies dominate the sample when number of
objects is considered.
The upper dashed lines
represent the cut used to separate red and blue galaxies,
as explained in \S4.1.
Since the evidence of color evolution in DEEP2 data is slight,
the zero-point and slope of this line with redshift is kept constant.
The lower dashed lines have the same slope and are used to divide
blue galaxies into two equal halves for further
luminosity-function analysis; their zero-points
are explained in \S4.2.

An interesting feature of these diagrams is that, even though the
detection of faint galaxies is favored at low redshifts, there
are still very few red galaxies found with $M_B>-18$, even at
redshifts below $z \sim$ 0.6, where they should be seen. The same
absence was also seen
by Weiner et al. (2005) in DEEP1 and by Kodama et al. (2004) in
distant clusters.  This point is discussed further
in the context of COMBO-17 data in Paper II.


\section{Methods}

\subsection{Luminosity Function Estimators}

The luminosity function is defined as the number of galaxies per unit
magnitude bin per unit co-moving volume,  and is most frequently
expressed using 
the Schechter (1976) parameterization, which in magnitudes is: 
\begin{eqnarray}
\phi(M) dM &=&0.4\; ln 10\; \phi^* 10^{0.4(M^*-M)(\alpha+1)}\nonumber \\ 
&{} & \times\; exp\lbrace -10^{0.4(M^*-M)} \rbrace dM,
\end{eqnarray}
where $\phi^*$  represents the characteristic number density of
galaxies per unit volume per unit magnitude, $M^*$ the characteristic magnitude where
the growth of the luminosity function changes from an exponential into
a power law, and $\alpha$ the slope of this power law that describes
the behavior of the faint end of this relation.
Several estimators have been proposed to measure this statistic (e.g.,
Schmidt 1968; Lynden-Bell 1971; Turner 1979; Sandage, Tammann \& Yahil
1979; Choloniewski 1986; Efstathiou, Ellis \& Peterson 1988), and the
relative merits of the different methods were explored
by Willmer (1997) and Takeuchi, Yoshikawa \&  Ishii (2000) through the
use of Monte-Carlo simulations. 

In this work, the luminosity function calculation relies on two
estimators. The first is the intuitive 
$1/V_{max}$ method where galaxies are counted within a volume.
The calculation  used here follows Eales (1993), Lilly et al. (1995),
Ellis et al. (1996) and Takeuchi et al. (2000), which overcomes the
bias identified by Felten (1976) and Willmer (1997).  The integral
luminosity function  for an absolute magnitude bin
between $M_{bright}$ and  $M_{faint}$ is described as:
\begin{equation}
\int_{M_{bright}}^{M_{faint}} \phi (M) dM =  \sum_{i=1}^{N_g} \frac {\chi_i}
{V_{max}(i)} ,
\end{equation}
where $\chi_i$ is the galaxy weight that corrects for the
sampling strategy used in the survey 
(discussed in detail in \S 3.3 below) and 
$V_{max}(i)$ is the maximum co-moving volume within
which a galaxy $i$ with absolute magnitude $M_i$ may be detected in the
survey:
\begin{equation}
V_{max}(i)  = \int _{\Omega} \int_{z_{min},i}^{z_{max},i} 
\frac {d^2V} {d \Omega dz} dz d \Omega,
\end{equation}
where $z$ is the redshift and $\Omega$ the solid angle being probed.
In a survey that is limited at bright ($m_l$) and
faint ($m_u$) apparent magnitudes, the redshift limits $z_{min},i$ and
$z_{max},i$ for galaxy $i$ are:
\begin{equation}
z_{max},i  = min\{z_{max}, z(M_i,m_u)\}
\end{equation}
\begin{equation}
z_{min},i  = max\{z_{min}, z(M_i,m_l)\}
\end{equation}
where the terms in braces
are the redshift limits imposed either by the limits of the redshift
bin being
considered ($z_{min}$ and $z_{max}$) or by the
apparent magnitude limits of the sample ($m_l$ and $m_u$). 
The Poisson error for the 1/$V_{max}$ 
method in a given redshift bin is given by:
\begin{equation}
\sigma_\phi = \sqrt {\frac {\chi_i}{(V_{max}(i))^2}}.
\end{equation}
In this paper, the $1/V_{max}$ method is calculated in absolute magnitude
bins 0.5 mag wide, and redshift bins of width $\Delta z$ = 0.2.
 The result is the average
value of the luminosity function $\phi(M_k,z)$ at redshift $z$
in magnitude bin $k$.  The method makes no assumption about the shape
of the luminosity function, therefore providing a non-parametric
description of the data.

The second estimator is the most commonly used in luminosity function calculations
-- the parametric maximum-likelihood method
of Sandage,  Tammann and Yahil (1979, STY; Efstathiou, Ellis \& Peterson 1988;
Marzke, Huchra \& Geller 1994).
The STY method fits an analytic Schechter function (Equation 2), yielding
values of the shape parameters $M^*$ and $\alpha$ (but not
the density normalization $\phi^*$).

The probability density that a galaxy with absolute magnitude $M_i$ will 
be found in a redshift survey sample is proportional to the ratio
between the differential luminosity function at $M_i$ and the
luminosity function integrated over the absolute magnitude range
that is detectable at redshift $z_i$.
In the case of DEEP2 galaxies, the STY conditional probabilities were modified
following Zucca, Pozzetti, \& Zamorani (1994) to account for the
galaxy weights, $\chi_i\ge1$ (see \S3.3 below), correcting for 
the sampling (e.g., Lin et al. 1999) and redshift success rates:
\begin{equation}
p(M_i,z_i) =\biggl\lbrack { \phi(M_i)dM \over \int_{M_{bright(z_i)}}^{M_{faint(z_i)}} \phi(M) dM }\biggl\rbrack^{\chi_i}.
\end{equation}
Here $M_{bright(z_i)}$ and $M_{faint(z_i)}$ are the absolute magnitude
limits at redshift $z_i$ accessible to a sample with apparent magnitude
limits $m_u$ and $m_l$.  $M_{bright(z_i)}$ and
$M_{faint(z_i)}$ are implicitly a function of
color (cf. Figure 4), which motivates the approach (used here) to
divide galaxies at least broadly into two color bins.
Implicitly, $\phi(M)$ is assumed to vary with $z$; the analysis
is carried out in fixed redshift bins in which $\phi(M,z)$ is
determined.

The likelihood function maximized by the STY method is defined by the
joint probability of all 
galaxies in the sample belonging to the same parent distribution. The
solution is obtained by assuming a parametric form for the luminosity
function and maximizing the logarithm of the likelihood function
relative to the product of the
probability densities of the individual galaxies $p(M_i,z_i)$:
\begin{equation}
\ln {\cal{ L}} = \ln \prod_{i=1}^{N_g} p(M_i,z_i).
\end{equation}

Because this method uses no type of binning, it preserves all
information contained in the sample. 
Since the luminosity function normalization is canceled out (Equation
8),
it is insensitive to density fluctuations in the
galaxy sample. However, this also means that the normalization (defined by
$\phi^*$) must be estimated separately, using the procedure described in
\S3.2 below.

Another shortcoming of the STY method is that it does not produce a
visual check of the fit. However, this can be done using the
$1/V_{max}$ method, which shows the average number density of
galaxies in bins of absolute magnitude and can be compared directly to the shape
parameters of the STY results. The $1/V_{max}$ points also provide an
independent check on the luminosity function normalization.


\subsection{ Luminosity Function Normalization}

Since the STY probability estimator is defined from the ratio
between the differential and integral luminosity functions, the density
normalization is factored out and has to be estimated
independently. The standard procedure for obtaining the luminosity
function normalization ($\phi^*$) measures the mean number
density of galaxies in the sample,  $\bar n$, which is then scaled  by the
integral of the luminosity function:
\begin{equation}
\phi^* = \frac { \bar n} { \int_{M_{\scriptscriptstyle
{bright}}}^{M_{\scriptscriptstyle{faint}}} \phi(M) dM }
\end{equation}
where $M_{bright}$ and $M_{faint}$ are the brightest and faintest absolute
magnitudes considered in the survey.
  
The method used to measure the mean density $\bar n$ is the unbiased
minimum-variance estimator proposed by Davis \& Huchra (1982):
\begin{equation}
\bar n = \frac {\displaystyle{ \sum_{i=1}^{N_g} \chi_i N_i(z_i) w(z_i)}}
{\int_{z_{min}}^{z_{max}} s(z) w(z) \frac {dV} {dz} dz },
\end{equation}
which averages the redshift distribution of galaxies,
$N_i(z_i)$, corrected by a weighting function, $ w(z_i)$, that takes
into account galaxy clustering; the selection function, $s(z$), that
corrects for the unobserved portion of the luminosity function; 
and the sampling weight, $\chi_i$.
The selection function is given by:
\begin{equation}
s(z) = \frac {\int_{max(M_{\scriptscriptstyle {min(z_i)}},M_{\scriptscriptstyle {bright}})
}^{min(M_{\scriptscriptstyle {max(z_i)}},M_{\scriptscriptstyle
{faint}})} \phi(M) 
dM} {\int_{M_{bright}}^{M_{faint}} \phi(M) dM}.
\end{equation}
where $M_{min(z_i)}$ and $M_{max(z_i)}$ are the brightest and faintest
absolute magnitudes at redshift $z_i$ contained within the apparent
magnitude limits of the sample.
The contribution due to galaxy clustering is accounted for
by the second moment $J_3$ of the two-point correlation function $\xi(r)$
(e.g., Davis \& Huchra 1982), which represents 
the mean number of galaxies in excess of random around each galaxy out
to a distance $r$ (typically set at $\sim$ 30 Mpc):
\begin{equation}
w(z_i) = \frac {1} { 1 + {\bar n} J_3 s(z)}, \quad J_3 = \int_0^r r^2
\xi(r) dr.
\end{equation}

Because of the small range of absolute magnitudes available at high
redshift, the shape of the faint end slope, parameterized by $\alpha$
is not constrained by the fit, so we opted to keep the value of this
parameter fixed, as discussed in \S4.2. Thus, in the calculation of
errors for the Schechter parameters only $M^*$ and $\phi^*$ are
considered.
Since the STY method factors out the density, it is also not
suitable for calculating the correlated 
errors of $\phi^*$ and $M^*$, as, lacking $\phi^*$,  
STY cannot take the high
correlation between these two errors into account.  
These errors were therefore calculated from the 1-$\sigma$ 
error ellipsoid (Press et al. 1992) that resulted from
fitting the Schechter function to the $1/V_{max}$
data points. Although the luminosity functions 
that result from the  STY
and $1/V_{max}$ methods are not quite identical (cf. Figure 7), 
the differences are small, and errors from $1/V_{max}$
should also be applicable to the STY method. 

\subsection{The Sampling Function and Galaxy Weights}

An issue with every data set is the selection of weights
to correct for missing galaxies.
The adopted weights need to take into account
the fact that 
(1) objects may be missing from the photometric catalog,
(2) stars may be identified as galaxies and vice versa,
(3) not all objects in the photometric catalog are targeted 
for redshifts (sampling rate)  and (4)
not all redshift targets yield successful redshifts (redshift success rate).
In the case of DEEP2, since the limiting magnitude of the  photometric
catalog is
1.5 magnitudes fainter than the limit adopted for redshift selection,
any effects due to incompleteness of the source catalog
should be negligible. The loss of galaxies brighter than
$R_{AB}$=24.1 but with surface brightness too low to admit them in the
photometric catalog  is ruled out from the
inspection of HST images analyzed by Simard et al. (2002) for the
EGS region in common with Groth Strip, which shows
no large low-surface brightness galaxies.
The loss of galaxies because of confusion with stars in well-defined
regions of the color-magnitude diagram (item 2 above) is shown in
\S3.4 below to be negligible. 
Therefore, only factors (3) and (4) need to be
taken into account in the weights.  The basic assumption to
deal with (3) is that all unobserved galaxies share the
same average properties as the observed ones in a given color-magnitude
bin.  The last effect, factor (4), is dealt with by assigning
a model redshift distribution to the failed galaxies.

A visual description of how the sampling rate and redshift success
rates depend on the magnitude and color of galaxies is shown in Figure
5, which projects both rates averaged in a color-color-magnitude data
cube onto the $R$ versus $R-I$ plane. Both rates are shown separately
for EGS and Fields 2-4 because of the different selection criteria.
In the EGS, the average sampling rate of slits placed on galaxies
is $\sim$60\%, and the average redshift success is 73\%.
For Fields 2, 3, and 4, the average sampling rate is 59\% (after
foreground galaxies are eliminated via color pre-selection),
and the redshift success is 73\%.

To account for the unobserved galaxies and redshift
failures we follow in this paper (and in Paper II for DEEP1 data) 
the method first applied by Lin et al. (1999) to the CNOC2 Redshift Survey.
This defines around each galaxy $i$  a data cube  in
color-color-magnitude space and, from all
attempted redshifts, counts the number of failed redshifts ($N_f$),
the number ($N_{z_h}$) of galaxies with $z>z_h$, where $z_h$ is the high
redshift limit of the sample;
the number ($N_{z_l}$) of galaxies with $z<z_l$, where $z_l$ is the low
redshift limit of the sample and
the number ($N_z$) with good redshifts within the ``legal'' redshift range
$z_l$ to $z_h$.  For Fields 2-4,  $z_l = 0.8$ and $z_h = 1.4$; and
for EGS these are  $z_l = 0.2$ and $z_h = 1.4$.
 
Next, for each galaxy in the photometric source catalog,
the probability that it has a redshift 
in the legal redshift range is    
estimated.  In the case of galaxies with good-quality redshifts, the
probability that the redshift lies in the legal range is
simply $P(z_l \leq z \leq z_h)$ = 1 when 
the galaxy has $z_l \leq z\leq z_h$ and
$P(z_l \leq z \leq z_h)$ = 0 for  $z >z_h$ or $z < z_l$.
To get the probability for unobserved galaxies, however,
some assumption must be made for the distribution of the {\it failed}
redshifts.  Two main models are
used in the present work. The first assumes that all
failed redshifts are beyond the high redshift cutoff of the sample, $z_h$
(the ``minimal'' model).
In this case, the probability that an unobserved galaxy
will be within the legal redshift range is the ratio of the number of
good redshifts in the
range divided by the sum of 
the number of successful redshifts plus failures:
\begin{equation} 
P(z_l \leq z \leq z_h) = {{ N_{z}} \over {N_{z}+N_{z_l}+N_{z_h}+N_f}}.
\end{equation}
The alternative model assumes that failures follow  
the same distribution as the observed sample
(the ``average'' model).  In this case, Equation 15 becomes:
\begin{equation} 
P(z_l \leq z \leq z_h) = {{ N_{z}} \over {N_{z}+N_{z_l}+N_{z_h}}}.
\end{equation}
Finally, the weight for each galaxy $i$ with an acceptable redshift is
calculated by adding for all galaxies $j$ within the color-color-magnitude
bin the probability that the redshift of galaxy  $j$ is within the
legal limits of the sample
\begin{equation}
\chi_i = { { {\sum_j P(z_l\leq z_j \leq z_h)}} \over {N_{z}}},
\end{equation}
where  $j$ includes both galaxies with and without attempted redshifts.

In the case of EGS, a final correction
is applied to the weights to account for the different
sampling strategy that was used, which includes low-redshift galaxies
but de-weights them so that they do not dominate the sample.
This (independently known) 
correction ($f_m$, Faber et al. in preparation; Newman et al. in
preparation) depends on the location of the galaxy in  
$B-R$ versus $R-I$ and its apparent magnitude. From this
correction, the probability that a galaxy will be placed on an EGS
mask is given by:
\begin{equation}
P(mask) =  0.33  +  0.43~P_{gal}~f_m,
\end{equation}
where $P_{gal}$ is the probability that an object is a galaxy. For
EGS galaxies the final probability weight is given by
\begin{equation}
\chi_i = { { {\sum_j P(z_{min} \leq z_j \leq z_{max})}} \over {N_{z}}~P(mask)},
\end{equation}
where  $j$ includes both galaxies with and without redshifts.

The comparison of Equations 14 and 15 shows that weights in the average
model are larger than in the minimal model.
The weights and differences in weights between the minimal and average models
are shown in Figure 6. These differences are  typically of order
15-20\% and  most large differences occur for galaxies
with extreme colors at faint magnitudes.

Based on the unpublished data of Steidel mentioned in \S 2, the minimal
model more closely matches blue galaxies 
since most failed blue galaxies lie beyond the upper redshift limit of
the survey $z_h$=1.4. In contrast, most failed red galaxies probably
lie within the survey range and are better described by the average
model. Because of this behavior, for the All galaxy sample
we adopt a compromise ``optimal'' model, where blue galaxies have
weights described by the minimal model, while red galaxies use the
average model. However, since the All sample is dominated by blue
galaxies, the differences between the optimal and minimal models are very small.

\subsection{Other Sources of Incompleteness}

Several tests were carried out to estimate the impact of what we
believe are the principal sources of  incompleteness,
namely
the surface brightness limit for slit
assignment, the misclassification of objects, 
and the presence of dropouts in the $B$ band photometry.

To limit the rate of redshift failures, a 
surface brightness cut (Equation 1) was used to  place galaxies
on slits. This restriction eliminates  both red ($R-I >$ 1.25) and blue ($R-I
\leq$ 1.25) galaxies, but the numbers are small.
The overall fraction of red galaxies that lie below the
surface brightness cut is $\sim$ 3\%, increasing to 6\% over the faintest
0.5 magnitude. For blue galaxies,
the average number is $\sim$ 5\%, increasing to 7\% in the faintest
0.5 magnitude bin. In both cases, these numbers are accounted for by
the weighting, since all galaxies that were not placed on slits are
still counted when the weights are calculated,  so no additional
corrections are needed, as long as  the 
characteristics of lower-surface brightness galaxies are similar to
those of other galaxies situated in the same color-color-magnitude bin.

As mentioned in \S2, the star-galaxy separation relies on the colors
and sizes of detected objects to assign each one the probability
of being a galaxy ($P_{gal}$).  Since stars occupy a well defined locus in the
$R-I$ $vs.$  $B-R$ diagram (Coil et al. 2004b), 
it is possible that DEEP2 galaxies with small apparent sizes and
observed colors close to the stellar locus could be treated as stars
($P_{gal} <$0.2) in DEEP2 mask-making and thus be ignored in the
analysis since the latter are not placed on masks.
This loss is estimated using objects in common between DEEP2 and the
structural catalog of Simard et al. (2002), which is derived from
psf-corrected photometry using HST images in the original Groth Strip.
When plotting the half-light radius $versus$ total-magnitude distribution
(e.g., Figure 6 of Im et al. 2002), stars and galaxies are well
separated down to the limiting magnitude $R_{AB}$= 24.1 
adopted by DEEP2, which corresponds to approximately $I814 \sim$ 23.5.
For red objects located close to the red stellar locus ($R-I \ge
1.25$, 1.8 $\leq B-R \leq$ 3.5), a total of 8 objects that are
clearly galaxies in HST images are identified as stars 
($P_{gal} <$0.2) in the DEEP2
source catalog, while 64 galaxies ($P_{gal} \geq $0.2) are correctly
identified within the same color boundaries. This corresponds to a 
loss of (8/64) or 13\%. 
On the other hand, the number of spectroscopically observed stars
misclassified as (red) galaxies, corresponds to $\sim$ 8\% of
the sample in the faintest magnitude bin ( 23.5 $\leq R_{AB} \leq$ 24.1).
An examination of the distribution of surface brightnesses shows
that all of these have $SB \leq$ 25 $R_{AB}$ $mag$ $arc~sec^{-2}$,
but that there are also galaxies in this range. Thus, the  inspection
of HST images and the distribution of sizes and surface brightnesses
suggests that DEEP2 may be biased against 
high surface brightness red galaxies with small
apparent sizes.  However, no strong dependence with redshift was seen.
Since the corrections for both effects are very uncertain, we opted not 
to apply them in the analysis. 

A final systematic error is caused by the presence of 
$B$-band dropouts, which are objects that have good $R$ and $I$
magnitudes, but a low S/N or non-existent $B$ measurement.
 All three magnitudes ($B, R, I$) are needed to sort
galaxies from stars; if $B$ is too dim and noisy, that object is
never assigned to a slit.  
Moreover, as there are no $B-R$ colors for the dropouts, such objects
are also not accounted for in the weighting procedure described above
which uses bins in color-color-magnitude space.
Consequently, the weights were modified  to account for the 
loss of these objects  by counting the number of dropouts within each 
($R$, $R-I$) bin around a given galaxy and dividing this number by
the total number of galaxies in the same bin. These corrections are
typically less than 4\%, though in some bins can reach $\sim$ 8\%, and are
applied to the final weights of each galaxy. The apparent $R-I$ colors
are consistent with most of these objects being part of the red
sequence. 

In summary, since most of these systematic effects due to
incompleteness are small, they will not affect the final conclusions
of this paper. Analyses carried out ignoring the last correction
produce essentially identical results to those in the present paper.

\section{Analysis}

\subsection{The Non-Parametric Luminosity Functions}

The DEEP2 luminosity function is shown in Figure 7, the top row
corresponding to the ``All'' galaxy function, while the second and
third rows show the luminosity function determined for sub-samples of
galaxies divided into ``Blue'' and ``Red'' by using the color
bimodality. The weighting model (\S 3.3) adopted for each population
is identified in the rightmost panel of each row.

For DEEP2 data, the color division between Red and Blue corresponds to
the upper dotted line in Figure 4, which is given by:
\begin{equation}
 U-B =  -0.032 (M_B + 21.52) + 0.454 -0.25.
\end{equation} 
This equation was derived from 
the van Dokkum et al. (2000) 
color-magnitude relation for red galaxies in distant clusters, 
converted to the cosmological model used in this paper and 
shifted downward by 0.25 mag in order to pass through the valley
between red and blue galaxies.  Although the colors of red galaxies
may evolve with redshift, this effect is not strongly seen in DEEP2
colors, and a line with constant zero-point independent of redshift
is adequate for all redshift bins. The constacy of $U-B$ constrasts
with the changes seen in the $U-V$  $vs.$ $M_V$ of COMBO-17 (B04).
However, when $U-B$, $U-V$ and $B-V$ colors are plotted as a function
of $z$  for the COMBO-17 sample, most of the color change can be traced to the
$B-V$ color (C. Wolf, private communication), implying that
the stability of the DEEP2 color-magnitude relation over this redshift
interval is not inconsistent with B04.

The separation between blue and red galaxies therefore is using a
clear feature which is easily identified, even if its physical
interpretation is not completely understood (e.g., Kauffmann et al. 2003).

Along the rows of Figure 7, each panel represents a different redshift bin,
with $z$ increasing from left to right.  
The DEEP2 non-parametric luminosity function estimated using the 1/$V_{max}$
method is represented by the solid black squares.
The sample used in the calculation of the luminosity function is shown
in Figure 4. The absolute magnitude range is truncated at the faintest absolute
magnitude which contains both red and blue galaxies, so that both
populations are sampled in an unbiased way. 
A fully volume-limited sample for a given
redshift bin would be 
obtained using the solid colored lines in Figure 4, which show
limiting absolute magnitudes of the upper redshift of each bin,
whereas the actually adopted limit (for the purpose of calculation of
the luminosity functions), corresponds to the lower redshift limit of
the bin. The slight loss of
galaxies in the remainder of the bin does not affect the STY estimation since
the range of absolute magnitudes accessible at any given $z$ is
calculated on a galaxy-by-galaxy basis. In contrast, the 
1/$V_{max}$ method will systematically underestimate the density of galaxies
unless corrected, which was done by following Page \& Carrera (2000),
The error bars 
represent counting errors assuming Poisson statistics only.  The
uncertainty due to cosmic variance is shown as a separate error bar at the top
left corner of each panel and was estimated following Newman \& Davis
(2002) who account for evolution of the correlation function using the
mass power spectrum, and using the correct field geometry, that takes
into account the elongated nature of DEEP2 fields which reduces the
cosmic variance . The
bias factors derived by 
Coil et al. (2004a) for red galaxies ($b = 1.32$) and blue galaxies ($b
= 0.93$) relative to the mass are included in these cosmic
variance estimates.  
To first order, cosmic variance should affect mainly
the overall number density, $\phi^*$, moving all points up and
down together and leaving the shape of the function unchanged,
whereas Poisson variance is random from point to point;
therefore we show the Poisson and cosmic variance
error bars separately.
The dashed gray curves represent the DEEP2 luminosity function fits (\S4.2)
measured in the lowest redshift bin ($0.2~\leq~z~<~0.4$), which are
repeated in subsequent panels. 
The major conclusions are as follows:

{\it All galaxies (top row):}  Relative to the low-$z$
Schechter function, the data in
successive redshift bins march to brighter magnitudes ($M^*_B$)
but remain roughly constant in number density ($\phi^*$).  
This visual assessment is confirmed by Schechter fits below.
In short, for the whole population,
galaxies are getting brighter with redshift, but
their number density is remaining much the same, to $z \sim 1$.

{\it Blue galaxies (middle row):} 
The results found above for the All sample
are repeated for the Blue sample, which is expected since
blue galaxies dominate the total number of galaxies.  This is shown in
the middle row of Figure 7. 
The increasing separation between the points and black solid lines in
each redshift  bin relative to the DEEP2 fits at
($0.2~\leq~z~<~0.4$) is easily seen, and  the visual impression is
that $M^*_B$ brightens
and $\phi^*$ remains constant, again confirmed by Schechter fits below.

{\it Red galaxies (bottom row):} 
The bottom row of Figure 7 presents the data for red galaxies.
As above, the dashed grey line represents 
the Schechter function fit to the lowest redshift bin of DEEP2 data.
In contrast to blue galaxies, between $z \sim$ 0.9 and $z \sim$ 0.3,
the luminosity function of red galaxies in DEEP2  shows no evidence 
for large changes, with most variations in the number density,
particularly at low $z$, being within the margins of cosmic variance.
The only bin that shows some hint of change is the highest-$z$ bin,
centered at $z $= 1.1, but which is likely to be the most affected by
incompleteness (see below).
Therefore the results from the DEEP2 survey alone are consistent with
rather little change in the raw counts of red galaxies at bright
magnitudes. If $M_B^*$ and $\phi^*$ are changing, they must do so in
coordinated fashion such that the counts at fixed magnitude remain roughly
constant. This behavior differs markedly from that of blue galaxies,
where counts increase at fixed $M_B$.

These results are fairly robust relative to the adopted weighting model.
The black points in Figure 7 use the average model of \S3.3, which
assumes that red galaxies without redshifts follow the same
distribution as the observed ones. 
For an extreme test, the weighting was changed to
a model 
where 
failed red galaxies (comprising about 25\%
of the total red galaxy sample) are {\it all placed in whatever
redshift bin is being considered}.  Here, red galaxies
are defined as all objects with
apparent $R-I>1.33$ (see line in Figure 1$b$).   
This extreme
assumption clearly yields a {\it strict upper limit}
to the red luminosity function in that bin.  The test works well
for red galaxies in the range
$ z = 0.7-1.1$, which all
cluster strongly near observed $R-I$ = 1.5 (see Figure 1$b$).

This part of the apparent CM diagram thus contains all
red galaxies that can possibly exist in this redshift range, unless
large numbers are missing from the photometric catalog,  which is unlikely, 
as discussed in \S3.4.
DEEP2 luminosity functions using
this extreme incompleteness model are shown
in Figure 8  as gray triangles.  It is important to note
that this model uses each failed red galaxy  multiple times
so the gray data points cannot be all valid simultaneously;
they are strict upper limits.
The new correction
does not increase the number of galaxies very much in the All function,
since the total counts are dominated by blue galaxies, and 
the Red function is significantly impacted in only the most
distant bin.  Quantitative conclusions are drawn below
by fitting Schechter functions.

\subsection {Schechter fits}

The Schechter functions fits using the STY method are presented here.
When splitting either galaxy sub-sample in narrow redshift bins, 
we see variations
in the best-fitting faint-end slope that are not statistically
significant, suggesting 
that we should average together slopes from several bins.  
In fact, the All galaxy function should show some trend
because the ratio of red to blue galaxies changes with redshift
and the shapes of the Red and Blue functions differ; however,
the effect is small.
As explained in more detail in Paper II, we decided to use the average 
faint-end slope values found within 
the range $z= 0.2$ to 0.6 for the COMBO-17 sample, because of the much
larger number of galaxies COMBO-17 contains in this redshift range 
in addition to there being no color pre-selection in that survey.
 The resulting  values of the faint-end slope are $\alpha=-0.5$ for the 
Red  sample and $\alpha=-1.3$ for the All and Blue samples;
these were applied also to DEEP2 here. Even though several recent works
have provided evidence of differential evolution
between bright and faint red galaxies (e.g., McIntosh et al. 2005;
Juneau et al. 2005; Treu et al. 2005), we adopt a fixed Schechter
function in shape at all redshifts. The effect of varying the shape is
small, as discussed in Paper II.
The evolving Schechter parameters are presented in Table 3 for the All sample
and in Tables 4 and 5 for the Blue and Red samples. 
Column (1) 
shows the central redshift of the bin; column (2) the number of
galaxies used in the luminosity function calculation in each redshift bin;
column (3) the value of the
adopted faint-end slope, $\alpha$; column (4) the value of $M^*_B$, followed by
the upper and lower 68\% Poisson errors in columns (5) and (6); the
mean density $\phi^*$ in column (7), followed by the 68\% Poisson errors 
in columns (8) and (9); the square root of the cosmic variance error
is shown in column (10); and (11) shows 
the luminosity density (in solar units) defined as
\begin{equation}
j_B(z) = {\int L \phi(L) dL} = L^*\phi^*\Gamma(\alpha+2),
\end{equation}
using $M_{B\sun}$=5.48 (Binney \& Merrifield 1998), where
$\Gamma$ is the Gamma function, with the 68\%
Poisson error in column (12); column (13)
indicates the weighting model (described in \S3.3)
used when calculating the fits. For the All and Red galaxy samples,
the results using the the upper-limit method of \S4.1 are also tabulated.
The 68\% Poisson errors for $M^*_B$ and $\phi^*$ were taken from the
$\Delta\chi^2$ = 1 contour levels in the ($M^*_B$, $\phi^*$) plane,
computed from the $1/V_{max}$ residuals and their errors relative to a
given Schechter fit. Cosmic variance errors were computed as described
above taking the volume and field geometry into
account and using separate bias ($b$) values for Blue and Red 
relative to the All galaxy sample.  Errors for $j_B$ were conservatively
calculated by adding the fractional Poisson errors for $M^*_B$ and
$\phi^*$ and cosmic variance in quadrature; these are an overestimate
because this neglects the correlated errors in $M^*_B$ and $\phi^*$,
which tend to conserve $j_B$. However, Poisson errors are generally
smaller than cosmic variance, which is dominant, so this overestimate is
small.

The changes of the Schechter parameters as a function of redshift are
shown in Figure 11 for $M^*_B$ (top row), $\phi^*$ (middle row) and
$j_B$ (bottom row) for the All, Blue and Red galaxy samples. The
figure shows results separately for minimal and average models, and in
the case of the All sample, using the optimal model. 
As expected from the raw counts in Figure 7, the Schechter parameters
for blue and red galaxies evolve differently with redshift. The
brightening of blue galaxies is clearly seen, while their number density
($\phi^*$) holds fairly steady. In contrast, red galaxies evolve only
modestly in either $M_B^*$ or $\phi^*$, and an increase in one
quantity is balanced by the other keeping the total  (red)
luminosity density, $j_B$, roughly constant out to the very last bin,
where it falls abruptly (see Table 5). The constancy of $j_B$ for
red galaxies was noted by Bell et al. (2004), who drew the conclusion
that the total stellar mass of the red sequence must be falling as a
function of increasing redshift. Paper II provides further evidence
for this. For now, we simply note that the DEEP2 red counts agree well
with the raw COMBO-17 red counts (as shown in Paper II), and with the
conclusion by Bell et al. (2004) that $j_B$ for red galaxies is
constant. 

The DEEP2 fitted values for $\phi^*$ also show a formally significant
drop back in time for red galaxies, a point which will be further
discussed in Paper II.

The DEEP2 data were also used to explore if the different trends measured
between red and blue galaxies can be detected when smaller
subdivisions in the color-magnitude space are considered. 
For this, blue galaxies were subdivided using a line parallel to Equation (19)
(which divides red from blue galaxies) but displaced downward 
in each redshift bin so it divides the blue galaxies into two equal halves.
This line was calculated considering only galaxies brighter than 
$M(z) = M_0 - Qz$, where $M_0$ = -20, and $Q$ is the amount of
luminosity evolution (measured in magnitudes) per unit redshift, so
that only the statistically similar  populations of galaxies would be used. 
This method was used in preference to a constant color cut,
which would yield a spurious evolution in numbers simply because blue 
galaxies are reddening with time (cf. Figure 4). 
Although this division does not use a clear feature as that dividing
blue and red galaxies, it is calculated at roughly the average
color of blue galaxies at a given absolute magnitude, and it allows
testing whether the degree of evolution is somehow correlated with the
average color of galaxies.
When calculating Schechter function fits for Moderately Blue and Very
Blue galaxies, we find that the fixed faint-end slope $\alpha$ = -1.3
used for the Blue galaxy sample provides a good description of both
sub-samples,
neither population shows significant
evidence that the faint-end slope is changing with redshift.
The evolution of $M^*_B$, $\phi^*$ and $j_B$ for the subsamples of
Moderately and Very
Blue galaxies is shown in Figure 10. The top row shows how $M^*_B$
changes with redshift, and it is readily apparent that the Moderately
Blue galaxies are on average more luminous than the Very blue
population.
On the other hand, the number density of both populations (second row)
does not show much evidence of significant changes; at all
redshifts, the Very Blue galaxies present higher number densities than
the Moderately Blue population. The luminosity density (bottom row) 
shows that, except for the highest redshift bin ($z \sim$ 1.3), 
Moderately Blue galaxies output most of the optical light coming from the
blue galaxy population. Overall both populations seem to evolve
similarly, maintaining a constant offset in $M^*_B$, while
$\phi^*$ holds constant for both halves separately. 
These results show that from $z \sim$ 1 to the present, most of the
light contributed by blue galaxies comes from galaxies with older stellar
populations and/or greater dust reddening than the typical star-forming
galaxy.

\section{Summary}

A sample of more than 11,000 DEEP2 galaxies from
$z=0.2$-1.4 is used to study
the evolution of galaxy luminosity functions.
When DEEP2 galaxies are plotted on the color-magnitude diagram ($M_B$
vs. $U-B$), blue and red galaxies occupy different loci, as seen in
local samples, and this 
division is still clearly seen at $z >$ 1.0.
The bimodality in the color-magnitude plane of galaxies is used to
subdivide the DEEP2 sample to study how luminosity functions evolve
as a function of galaxy color. 
In order to account for the partial sampling strategy and
redshift success rate of DEEP2 as a function of color and magnitude,
weights are calculated using different models describing how
failed redshifts are distributed in $z$.
The current data suggest that the vast majority of faint and blue
galaxies in the DEEP2 sample for which no redshifts were successfully
measured are at high redshift ($z > 1.4$). In this work we make the
assumption that red galaxies with failed $z$'s follow roughly the same
redshift distribution as the good measurements.
Given the nature of redshift failures, a compromise approach where 
blue failures are assumed to be at high redshift (minimal), while red
failures are assumed to follow the average model is regarded as optimal.
The conclusions of this work hold independently of the adopted model.

The results from this work show that populations of blue and red
galaxies evolve 
differently. As an ensemble, blue galaxies show a larger amount of
luminosity evolution, yet show little change in overall number density.
Red galaxies show less change in luminosity, but a larger change
in number density. When the luminosity density is considered, blue
galaxies show a steady decrease toward lower redshifts, while the
luminosity density of red galaxies is almost constant. 

Finally, we divided the blue galaxies using the a sloping line that splits
the population into two equal halves at each redshift.
We find that both halves are still adequately described
by a fixed faint end slope of $\alpha$= -1.3, and that both
sub-populations evolve in a similar manner.  Even in our
highest redshift bins, the adopted shape of the faint end still
provides a good description of the data, with no strong
evidence of an increase in numbers of Very Blue galaxies at the lowest
luminosity limit we probe. 

A detailed comparison between the results obtained for the DEEP2
survey (this paper) with other works (Wolf et al. 2003; Bell et
al. 2004; Gabasch et al. 2004; Ilbert et al. 2005) shows a good
agreement. The combined results of these surveys are presented in
Paper II (Faber et al. 2005), suggesting that the luminosity function
of galaxies to $z \sim$ 1, is currently well understood.
The present paper presents the results using about a quarter of the 
planned DEEP2 data, and shows the potential that DEEP2 has in
characterizing the properties of galaxy populations to $z \sim$ 1.2.
As the DEEP2 survey reaches completion, ancillary data coming from
$Z$-band photometry by Lin and collaborators are also being obtained
in DEEP2 Fields 2-4. These will allow measuring
photometric redshifts for galaxies in these three fields, and will
allow a far more precise characterization of the properties of
galaxies with ``failed'' redshifts. This, combined with a 4 $\times$
larger sample with spectroscopic redshifts, will constitute for many
years to come the main sample of galaxies at redshifts 0.7 $\leq z
\leq$ 1.4, that can be used to study how galaxy populations change with time.
The data for Fields 2-4 used in this paper can be retrieved from
{\it{http//deep.berkeley.edu/DR1}}. The second data release,
tentatively scheduled for late 2005, will included all the data which
were used in the analysis of this paper.

\acknowledgments

The DEEP team thanks C. Wolf for several discussions regarding the
color-separated luminosity function and C. Steidel for sharing
unpublished redshift data. 
CNAW thanks G. Galaz, S. Rauzy, M. A. Hendry and K. D'Mellow for
extensive discussions on the measurement of the luminosity function.
Suggestions from the anonymous referee are gratefully acknowledged.
The authors thank the Keck Observatory
staff for their constant support 
during the several observing runs of DEEP2; the W. M. 
Keck Foundation and NASA for construction of the Keck telescopes.
The DEIMOS spectrograph was funded by NSF grant ARI92-14621 and by
generous grants from the California Association for Research in
Astronomy, and from UCO/Lick Observatory.
We also wish to recognize and acknowledge the highly significant
cultural role and reverence that the summit of Mauna Kea has always 
had within the indigenous Hawaiian community. It is a privilege to be
given the opportunity to conduct observations from this mountain. 
Support from National Science Foundation grants 00-71198 to UCSC and
AST~00-71048  to UCB is gratefully acknowledged.
SMF would like to thank the California
Association for Research in Astronomy for a generous
research grant and the Miller Institute at UC Berkeley for the support
of a visiting Miller Professorship.
JAN acknowledges support by NASA through Hubble Fellowship grant 
HST-HF-01132.01 awarded by the Space Telescope Science Institute, which 
is operated by AURA Inc. under NASA contract NAS 5-26555.
 Computer hardware gifts from Sun Microsystems and
Quantum, Inc. are gratefully acknowledged.  
This research has made use of the NASA/IPAC Extragalactic Database (NED),
which is operated by the Jet Propulsion Laboratory, California
Institute of Technology, under contract with the National Aeronautics
and Space Administration. Finally, we acknowledge NASA's 
(indispensable) Astrophysics Data System Bibliographic Services.

\clearpage

\begin{figure}
\vspace{80mm}
\includegraphics{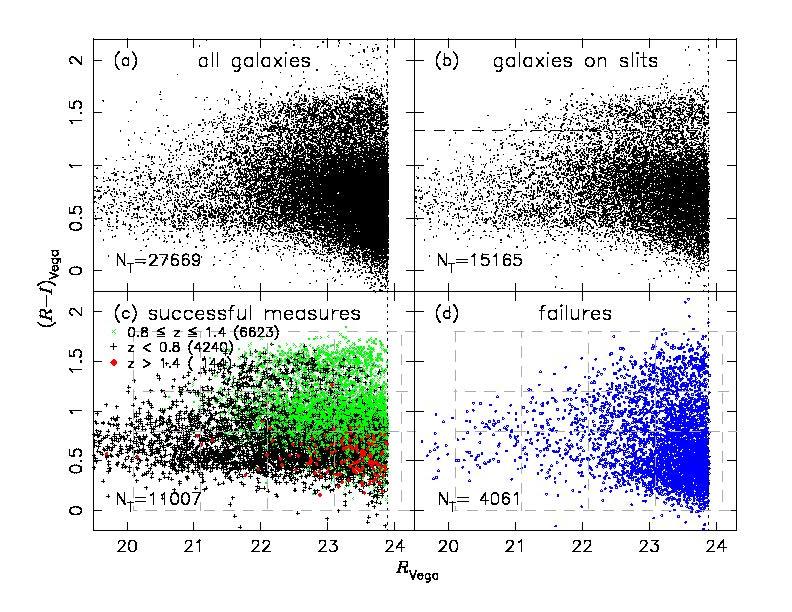}
Figure 1 is available as 0506041.f1.jpg
\caption{Apparent color-magnitude distribution of galaxies in the
  DEEP2 Survey. Panel $a$ shows the full sample,
  panel $b$ the distribution of galaxies set on slits, and
  panel $c$ the distribution of successful redshifts, where
  galaxies in the main DEEP2 redshift interval are shown as green
  crosses, galaxies lying beyond
  the upper redshift limit adopted in this work ($z =$ 1.4) are shown as red
  diamonds, and galaxies below the main redshift limit adopted here ($z =$ 0.8)
  are black plusses. Panel $d$ shows the
  distribution of failed redshifts. 
  The $R_{AB}$ limiting
  magnitude of 24.1, 
  transformed into $R_{Vega} = 23.88$, is shown as the
  vertical dotted line. The ridge
  of galaxies at blue colors ($R-I\sim 0.5$) is dominated by galaxies
  at redshifts 
  below the DEEP2 pre-selection color cut at $z = 0.7$; faint ones also
  include many distant galaxies with $z > 1.4$. The bimodal
  distribution seen in rest-frame colors (cf. Figure 4) is also seen in
  observed $R-I$; the horizontal dashed line in panel $b$ shows the dividing
  line for the extreme red-galaxy correction function used in
  \S4.1 at $R-I$ = 1.33. Dashed
  grey lines show the boundaries used in the redshift histograms
  displayed in Figure 2.
}
\end{figure}

\begin{figure}
\vspace{90mm}
\includegraphics{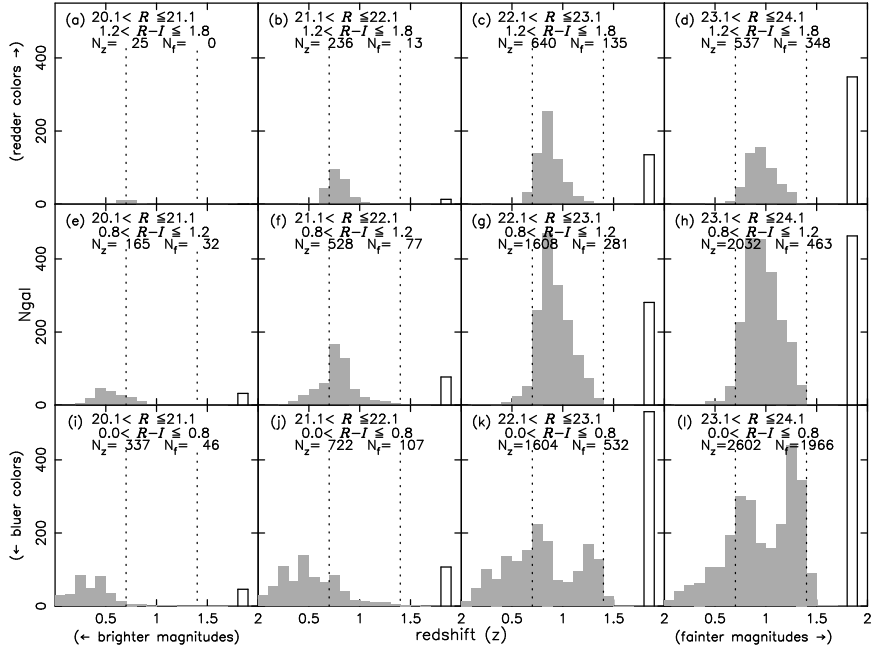}
\caption{Raw redshift distributions of DEEP2 galaxies in the apparent
  color-magnitude bins indicated in Figure 1.  Magnitudes
  become fainter towards the right and colors redder towards the top.
  The color boundaries were chosen to correspond
  roughly to the main loci of galaxies in rest-frame
  color-magnitude space:
  the top four panels correspond to
  galaxies on the red sequence, while the middle panels
  correspond to intermediate colors, i.e., $(U-B) \le -0.3$
  mag bluer than the red sequence, as outlined by the middle
  dotted line in Figure 4. The lower panels represent galaxies 
  bluer than this.
  The vertical dashed lines represent the low- and high-$z$ 
  design limits for Fields 2, 3,
  and 4 of DEEP2 ($z$ = 0.7 and 1.4 respectively); galaxies
  at lower redshifts come mainly from Field 1 (EGS). No attempt is
  made in this plot to correct for the different slit assignment
  algorithm used for EGS and Fields 2-4.
  Failed redshifts are represented by the bars to the right of each
  panel; the one at lower right has been truncated to 550 galaxies.
  The total number of galaxies plotted is shown in each panel, where
  $N_z$ represents the number of good measurements and $N_f$ the
  number of failures. 
  The bimodal distribution in redshift seen in the two fainter
  magnitude bins for blue galaxies is an artifact caused by the shift
  of the 4000 \AA~ break into and out of the $R$ and $I$ filters as a
  function of redshift.
  When both filters are redder than 4000 \AA~ the colors are flat,
  then become red, then flatten again once both filters are bluer than
  rest-frame 4000 \AA.
}
\end{figure}

\begin{figure}
\vspace{80mm}
Figure 3 is available as 0506041.f3.jpg
\includegraphics{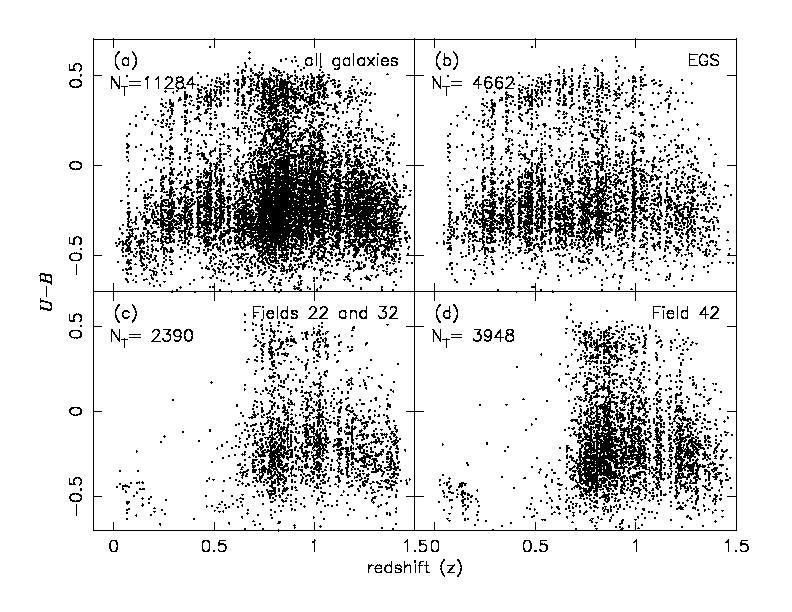}
\caption{Rest-frame $U-B$ as a function of redshift 
  for DEEP2 galaxies. The $U-B$ values in this paper are corrected for
  Galactic extinction but not for internal galactic extinction.
  The bimodal distribution of colors is clearly seen to $z \sim$ 1. 
  Fields 2, 3, and 4 lack
  low-redshift objects because of the pre-selection
  color cut; this cut was not
  applied to Field 1 (EGS), but a secondary redshift selection still
  applies to this field, as explained in the text.  The lack of low-redshift 
  red galaxies (in EGS) is likely due the combination of 
  the relative paucity of red galaxies, the small volume at low
  redshifts, and the bright apparent magnitude cut of the sample.
  All fields show
  clustering, and the observed variations are due to cosmic variance.
}
\end{figure}
\begin{figure}
\vspace{90mm}
Figure 4 is available as 0506041.f4.jpg
\includegraphics{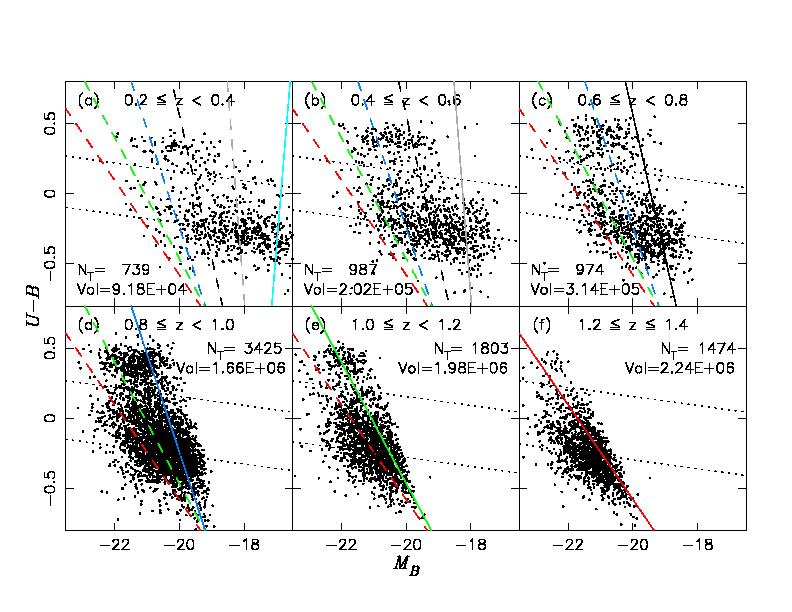}
\caption{Rest-frame color-magnitude diagrams for DEEP2 for the
  redshift intervals used in this work. 
  The three lower $z$ intervals (panels $a$, $b$ and $c$) 
  contain only EGS data, while galaxies in all four DEEP2
  fields are shown for $z\ge 0.8$.
  The solid line in each panel indicates
  the approximate faint absolute magnitude limit as a function of
  intrinsic color and redshift for a sample with a fixed apparent
  magnitude limit $R_{Vega}=23.88$. This line is calculated at the $upper$
  redshift limit of each panel and denotes the limiting magnitude
  for which a volume-limited sample could be defined in
  that bin.
  This calculation uses the distance modulus and the K-correction
  appropriate for each template SED, which is then fit by a linear
  relation, corresponding to the plotted line.
  The dashed lines repeat the same lines in other panels.
  The upper dotted
  line denotes the cut used to define red-sequence galaxies
  (Equation 19) and is the same at all redshifts.  
  The lower dotted line is drawn parallel
  to this, but its vertical height is displaced downward in each 
  redshift bin to divide Very Blue from Moderately Blue
  galaxies into two equal halves (see \S4.2). 
  The numbers in
  each panel show the number of galaxies
  plotted and the co-moving volume in {\rm{Mpc}}$^{3}$ for the ($H_0, \Omega,
  \Lambda)$ = (70, 0.3, 0.7) cosmology.
}
\end{figure}
\begin{figure}
\vspace{80mm}
\includegraphics{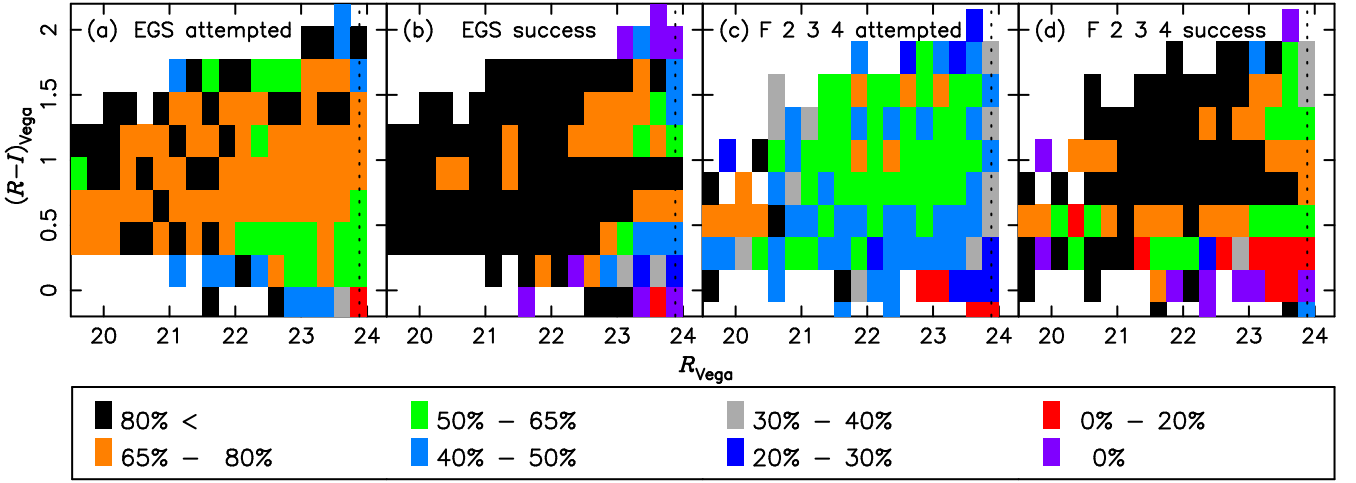}
\caption{Sampling and redshift success rates 
  as a function of apparent magnitude and $R-I$ color for DEEP2. The
  colors, coded in the key, correspond to the
  sampling rate (panels $a$ and $c$) or redshift success rate (panels
  $b$ and $d$).
  In all panels, the black dotted line corresponds
  to the limiting apparent magnitude $R_{Vega}=23.88$.  Panels $a$ and $b$
  refer to the Extended Groth Strip, while panels $c$ and $d$ 
  show Fields 2, 3, and 4.  Panels $a$ and $c$ show
  the percentage of galaxies placed on slits relative to the total sample
  (for EGS, the total sample is all galaxies in each $R, R-I$ bin;
  for Fields 2, 3, and 4, it is the target galaxies
  photometrically selected to have $> 0.7$).  Panels $b$ and $d$
  show the success rate for good redshifts of those
  attempted. The difference in 
  sampling rates between the EGS and Fields 2-4 is caused by
  differences in the density of targets and slitmasks on
  the sky for these fields, along with changes to the  weights given
  faint objects made after the early data from Fields 2-4  were obtained.
}
\end{figure}
\begin{figure}
\vspace{90mm}
Figure 6 is available as 0506041.f6.jpg
\includegraphics{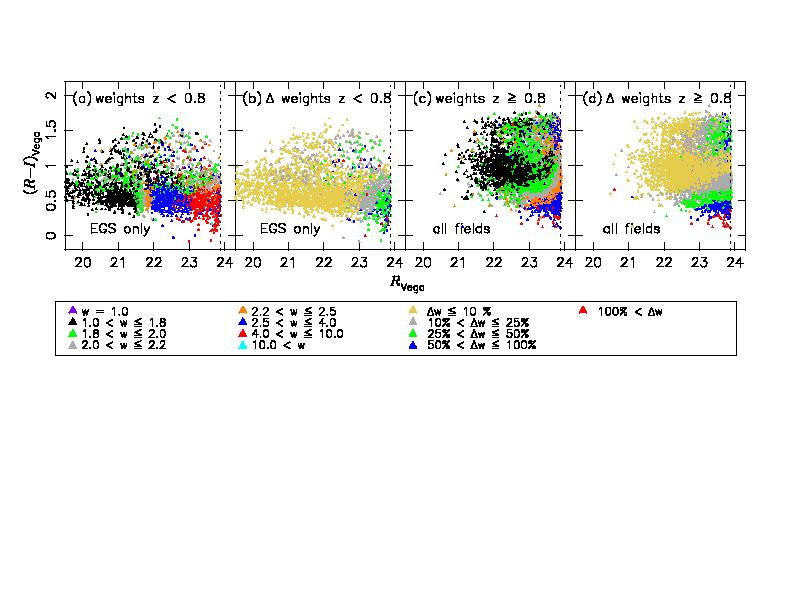}
\caption{Weights used to correct
  for incomplete sampling and failed redshifts.
  Panels $a$ and $b$ show
  galaxies in EGS with $z < 0.8$, while panels $c$ and $d$
  show galaxies in all
  four fields with $z \ge 0.8$. 
  Weights are shown here as a function of $R~vs.~R-I$. In actuality, they are
  calculated in bins of color-color-magnitude space, incorporating $B-R$ as well.
  Panels $a$ and $c$ show the weight of each galaxy
  using the ``minimal'' model, in which all
  galaxies with failed redshifts are  assumed to lie
  above the upper redshift limit of the survey
  ($z=1.4$). 
  Panels $b$ and $d$ show how the galaxy weights change in moving from the
  minimal model to the ``average'' model, in which failed galaxies are
  assumed to be distributed in $z$ like the observed ones. 
}
\end{figure}
\begin{figure}
\vspace{110mm}
\includegraphics{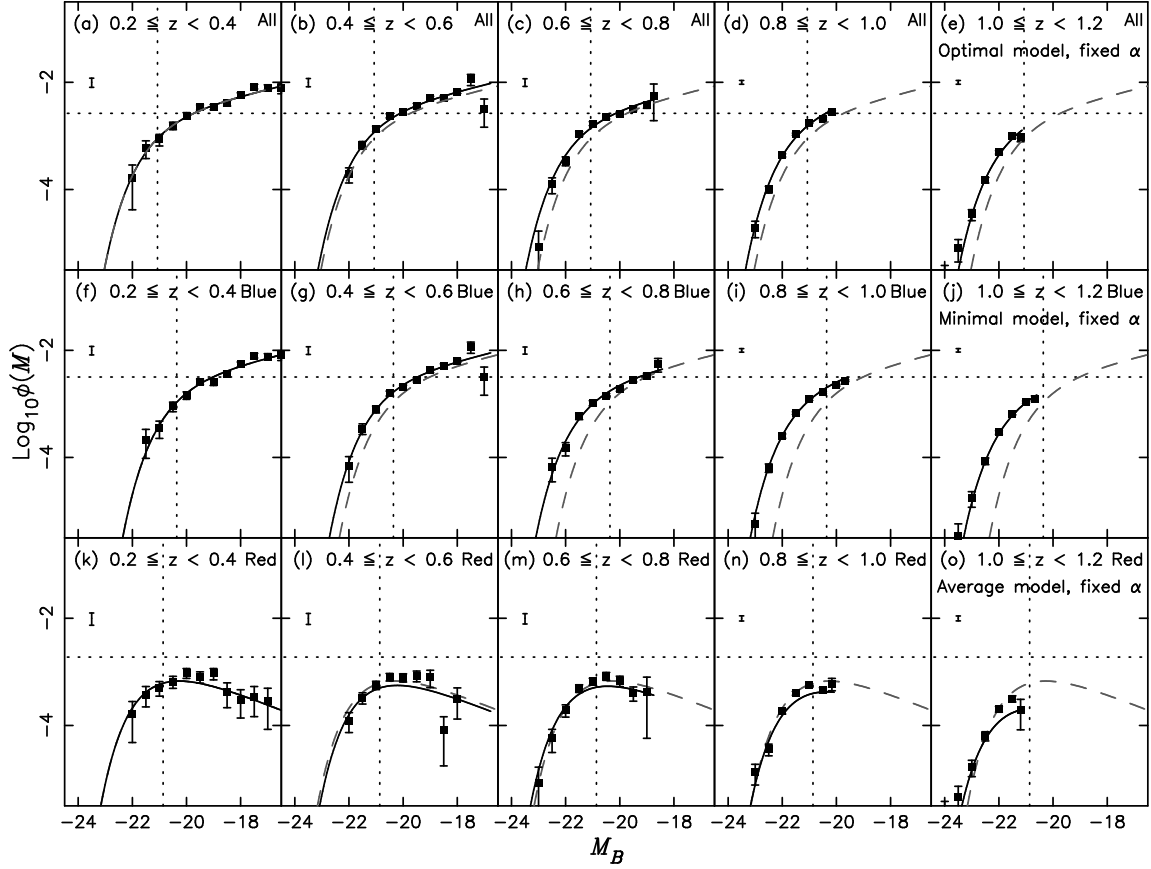}
\caption{Luminosity functions measured in different redshift bins for
  ``All'' galaxies (top row), ``Blue'' galaxies (middle row), 
  and ``Red'' galaxies (bottom row).  Points calculated using the
  $1/V_{max}$ method are shown as black squares.
  Error bars represent the 68\% Poisson error bars only. Errors due to
  cosmic variance (calculated as described in the text) are shown at
  the top left of each panel.
  The values plotted use the favored models to
  correct for the incomplete sampling rate and redshift failures (see text).
  The solid black lines represent the STY
  fits to  DEEP2 data, keeping the faint-end slope
  $\alpha$'s fixed at the values measured
  from the COMBO-17 ``quasi-local'' sample in three redshift bins of
  $\Delta z$ = 0.2 width, ranging from $z = 0.2-0.6$ (see
  Paper II).
  The values assumed  are $\alpha = -1.3$ (All), $\alpha = -1.3$ (Blue),
  and $\alpha = -0.5$ (Red). The dashed grey curves show the Schechter
  function fits to the lowest redshift bin measured by DEEP2 and are
  repeated in each panel for All, Blue and Red galaxies respectively.
  The dotted lines serve as a visual reference
  and are plotted at the values of  $M^*_B$ and $\phi^*$ for the
  lowest redshift interval. The main conclusion from this figure is
  that blue and red luminosity functions evolve differently: blue
  counts at fixed absolute magnitude increase markedly back in time,
  while red counts tend to remain constant.
}
\end{figure}
\begin{figure}
\vspace{60mm}
\includegraphics{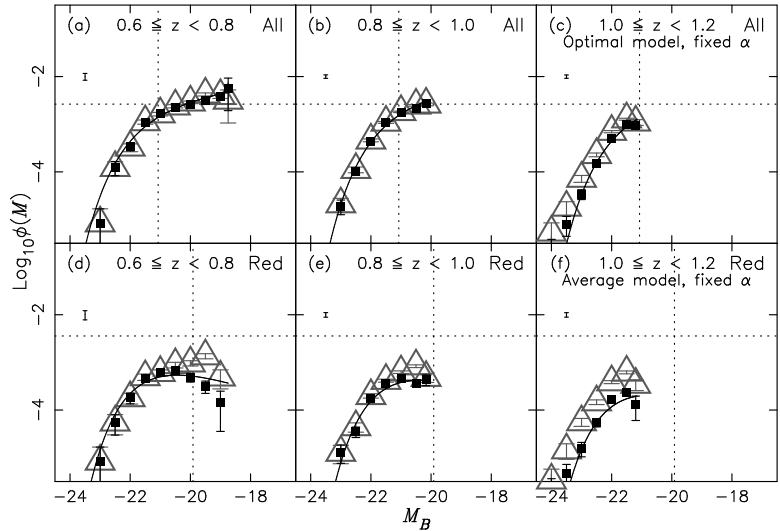}
\caption{Luminosity function for DEEP2 All and Red galaxy
  samples. The change relative to Figure 7 is the addition of the grey
  triangles, which are strict upper limits to the density of galaxies 
  under the
  extreme assumption that all failed-redshift red galaxies are
  located in that bin only.  This model uses each failed red galaxy
  more than once, and thus all grey triangles cannot be valid
  simultaneously.  Only in the highest
  redshift bin does the use of this assumption cause a significant
  increase in the number density of red galaxies.  The solid black lines show
  the same parametric fits as in Figure 7. 
}
\end{figure}
\begin{figure}
\vspace{100mm}
\includegraphics{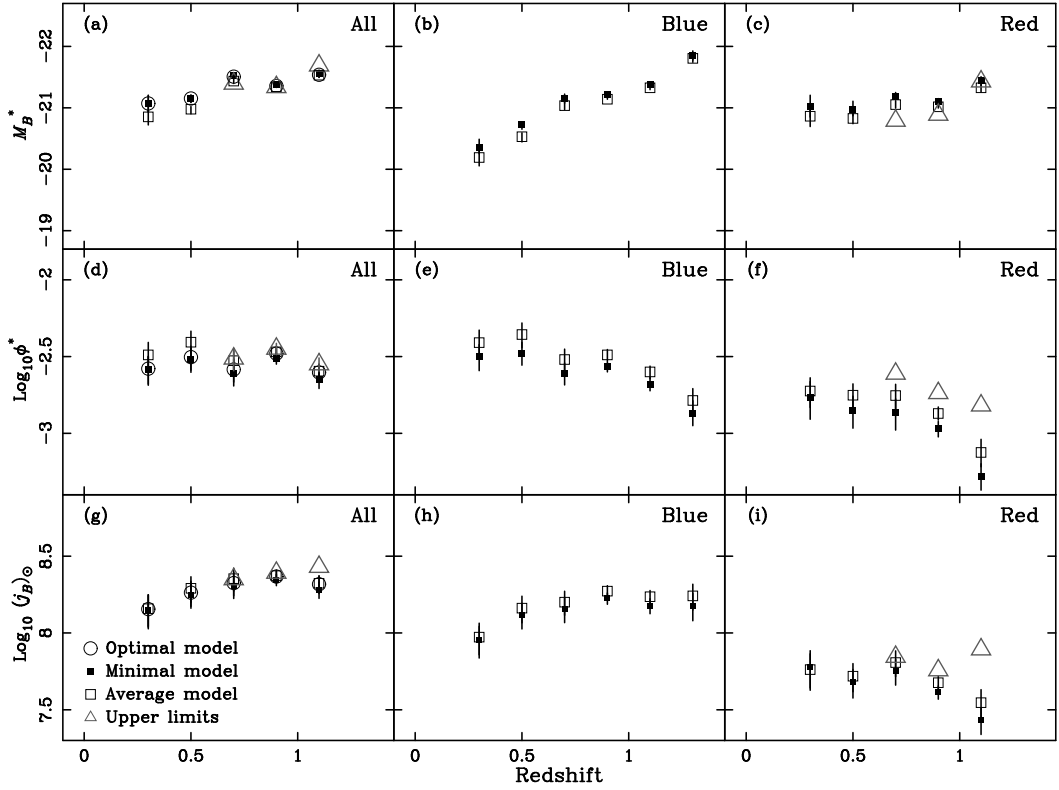}
\caption{Evolution of the Schechter function parameters 
  $M^*_B$,  $\phi^*$ and $j_B$
  assuming constant $\alpha$, as a function of redshift. 
  The solid black squares show the values measured using the minimal
  weight model, open squares show the values for the average
  model, and open circles the optimal model. Because blue galaxies
  dominate the total numbers of galaxies, the results using optimal and minimal
  weights are very similar. The minimal model is prefered for Blue, while the
  average model is preferred for Red galaxies (see text).
  The open triangles 
  represent the fit values making the extreme assumption that all
  red galaxies for which no redshift could be measured are located in the 
  $z$ = 0.7, 0.9 and 1.1 bins respectively, providing absolute upper limits
  for the Schechter parameters.  The difference in the mode of
  evolution for blue and red galaxies is clearly seen. The quantity
  $M_B^*$ increases markedly back in time for blue galaxies, while number density
  $\phi^*$ holds roughly constant (to $z=1$). The net effect is that
  $j_B$ for blue galaxies is increasing with redshift. Magnitude
  evolution of red 
  galaxies is milder, though $\phi^*$ may drop
  more. The net effect is that $j_B$ for red galaxies remains relatively
  constant than for blue galaxies to $z = $0.9, but may drop beyond that.
}
\end{figure}
\begin{figure}
\vspace{70mm}
\includegraphics{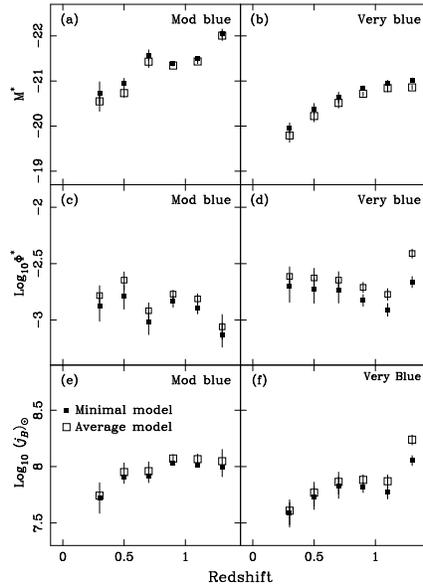}
\caption{Evolution of the Schechter function parameters
  $M^*_B$,  $\phi^*$ and $j_B$
  assuming constant $\alpha$, as a function of redshift for Moderately
  Blue and Very Blue galaxies.
  The solid black squares show the values measured using the minimal
  weight model, while open squares show the values for the average
  model. On average, the Moderately Blue galaxies are brighter than
  the Very Blue galaxies, both populations showing a comparable
  brightening back in time. The number density of Very Blue galaxies
  is always slightly higher than of Moderately Blue galaxies, and both
  populations show small changes back in time. Both Moderately Blue
  and Very Blue galaxies show a steady increase in luminosity density
  as higher redshifts are reached. Up to $z\sim$ 1.1, the $j_B$
  measurements suggest that the bulk of the light from blue galaxies
  comes from the Moderately Blue population.  Overall the properties
  of both populations of galaxies show a similar evolutionary trend.
  These results from $j_B$ suggest that to redshifts reached by DEEP2, 
  the optical light from galaxies is not dominated by newly-formed 
  stars but rather by a combination of these with older stellar
  populations.
}
\end{figure}

\clearpage
\begin{table}
\caption{Conversion between AB and Vega Magnitudes}
\begin{center}
\begin{tabular}{llll}
\hline
\hline
{} & {} & Transformation & System \cr
\noalign{\smallskip} \hline \noalign{\smallskip}
$U_{AB}$     &   =  & $U_{Vega}$    + 0.73  & Johnson  \cr
$B_{AB}$     &   =  & $B_{Vega}$   -- 0.10  & Johnson  \cr
$(U-B)_{AB}$ &   =  & $(U-B)_{Vega}$ + 0.81 & Johnson  \cr
$B_{AB}$     &   =  & $B_{Vega}$   -- 0.11  & CFHT 12K$\times$8K B \cr
$R_{AB}$     &   =  & $R_{Vega}$    + 0.22  & CFHT 12K$\times$8K R \cr
$I_{AB}$     &   =  & $I_{Vega}$    + 0.44  & CFHT 12K$\times$8K I \cr
$(B-R)_{AB}$ &   =  & $(B-R)_{Vega}$  -- 0.33  & CFHT 12K$\times$8K \cr
$(R-I)_{AB}$ &   =  & $(R-I)_{Vega}$  -- 0.22  & CFHT 12K$\times$8K \cr
\noalign{\smallskip} \hline
\end{tabular}
\end{center}
\ \ Note. -- 
The procedure used to calculate these transformations is
described in Appendix A
\end{table}

\begin{table}
\caption{Survey Characteristics}
\begin{center}
\begin{tabular}{lrrrrrrrrrl}
\hline
\hline
Survey            &
Area               &
$N_{field}$       &
$N_{gal}$         &
$N_z$             &
$N_z~>~0.8$       &
$m_l$             &
$m_u$             &
$z_{min}$         &
$z_{max}$         &
System            \\
{}                &
$\sq\degr$        &
{}                &
{}                &
{}                &
{}                &
{}                &
{}                &
{}                &
{}                \\
(1)               &
(2)               &
(3)               &
(4)               &
(5)               &
(6)               &
(7)               &
(8)               &
(9)               &
(10)              &
(11)              \\
\noalign{\smallskip} \hline \noalign{\smallskip}
EGS           & 0.28 &1 & 9115  & 4946 & 2026 &18.5 & 24.1 & 0.2 & 1.4 & $R_{AB}$   \\
Fields 2+3+4  & 0.85 &3 & 18756 & 6338 & 4820 &18.5 & 24.1 & 0.8 & 1.4 & $R_{AB}$   \\
\noalign{\smallskip} \hline
\end{tabular}
\end{center}
\ \ Note. --
The meanings of columns are:  (1) Surveyed region;
(2) area in square degrees; 
(3) number of non-contiguous fields in  surveyed region;
(4) number of galaxies in source catalogue;
(5) number of good quality redshifts;
(6) number of good quality redshifts above $z$ = 0.8;
(7) bright apparent magnitude limit;
(8) faint apparent magnitude limit;
(9) lower redshift limit;
(10) upper redshift limit;
(11) apparent magnitude system of catalogue.
\end{table}

\begin{table}
\caption{Schechter function parameters for All galaxy samples}
\begin{center}
\begin{tabular}{crcrrrrrrrrcl}
\hline
\hline
$\langle z \rangle$ & N$_{gal}$ & $\alpha$ &\multicolumn{3}{c}{$M_B^*$} &\multicolumn{3}{c}{$\phi^*$} &$\sqrt{Var}$&\multicolumn{2}{c}{$j_B$} & Weights \cr
 {} & {} & {} & \multicolumn{3}{c}{ } &\multicolumn{3}{c}{$\times$ 10$^{-4}$ Gal Mpc$^{-3}$} & {} &\multicolumn{2}{c}{$\times$ 10$^{8}$ $L_{\odot}$}&{} \cr
 (1) & (2) & (3) & (4) & (5) & (6) & (7) & (8) & (9) & (10) & (11) & (12) & (13)\cr
\noalign{\smallskip} \hline \noalign{\smallskip}
      0.30 &   734 &  -1.30 &  -21.07 & (+   0.13 & -   0.14) &   26.08 &(+    2.10 & -   2.12) &    0.20 &    1.41 & $\pm$   0.35 & minimal         \cr
      0.50 &   983 &  -1.30 &  -21.16 & (+   0.05 & -   0.07) &   30.40 &(+    0.97 & -   0.98) &    0.18 &    1.78 & $\pm$   0.33 & minimal         \cr
      0.70 &   914 &  -1.30 &  -21.53 & (+   0.03 & -   0.03) &   24.43 &(+    0.75 & -   0.80) &    0.16 &    2.02 & $\pm$   0.34 & minimal         \cr
      0.90 &  2561 &  -1.30 &  -21.38 & (+   0.01 & -   0.02) &   30.81 &(+    0.30 & -   0.63) &    0.08 &    2.22 & $\pm$   0.19 & minimal         \cr
      1.10 &   844 &  -1.30 &  -21.57 & (+   0.05 & -   0.04) &   22.36 &(+    2.09 & -   1.39) &    0.08 &    1.91 & $\pm$   0.23 & minimal         \cr
{} &{} &{} &{} &{} &{} &{} &{} &{} &{} &{} &{} \cr
      0.30 &   740 &  -1.30 &  -20.85 & (+   0.10 & -   0.13) &   32.46 &(+    1.88 & -   2.14) &    0.20 &    1.44 & $\pm$   0.34 & average         \cr
      0.50 &   983 &  -1.30 &  -20.98 & (+   0.04 & -   0.08) &   39.26 &(+    1.94 & -   1.65) &    0.18 &    1.95 & $\pm$   0.37 & average         \cr
      0.70 &   919 &  -1.30 &  -21.44 & (+   0.04 & -   0.04) &   29.77 &(+    0.99 & -   0.90) &    0.16 &    2.26 & $\pm$   0.39 & average         \cr
      0.90 &  2436 &  -1.30 &  -21.34 & (+   0.03 & -   0.01) &   33.98 &(+    0.60 & -   0.39) &    0.08 &    2.37 & $\pm$   0.20 & average         \cr
      1.10 &   805 &  -1.30 &  -21.53 & (+   0.04 & -   0.03) &   25.43 &(+    2.20 & -   1.88) &    0.08 &    2.11 & $\pm$   0.25 & average         \cr
{} &{} &{} &{} &{} &{} &{} &{} &{} &{} &{} &{} \cr
      0.30 &   734 &  -1.30 &  -21.07 & (+   0.13 & -   0.13) &   26.39 &(+    1.81 & -   1.62) &    0.20 &    1.43 & $\pm$   0.33 & optimal         \cr
      0.50 &   983 &  -1.30 &  -21.15 & (+   0.06 & -   0.06) &   31.39 &(+    0.97 & -   1.04) &    0.18 &    1.83 & $\pm$   0.32 & optimal         \cr
      0.70 &   914 &  -1.30 &  -21.51 & (+   0.03 & -   0.03) &   26.07 &(+    1.39 & -   1.14) &    0.16 &    2.11 & $\pm$   0.34 & optimal         \cr
      0.90 &  2561 &  -1.30 &  -21.36 & (+   0.01 & -   0.02) &   33.04 &(+    0.90 & -   1.11) &    0.08 &    2.33 & $\pm$   0.20 & optimal         \cr
      1.10 &   844 &  -1.30 &  -21.54 & (+   0.04 & -   0.04) &   24.94 &(+    2.20 & -   2.63) &    0.08 &    2.08 & $\pm$   0.27 & optimal         \cr
{} &{} &{} &{} &{} &{} &{} &{} &{} &{} &{} &{} \cr
      0.70 &  1059 &  -1.30 &  -21.39 & (+   0.04 & -   0.05) &   30.70 &(+    0.86 & -   1.08) &    0.16 &    2.24 & $\pm$   0.42 & upper limit     \cr
      0.90 &  2844 &  -1.30 &  -21.34 & (+   0.01 & -   0.01) &   35.60 &(+    0.83 & -   0.22) &    0.08 &    2.47 & $\pm$   0.40 & upper limit     \cr
      1.10 &  1210 &  -1.30 &  -21.69 & (+   0.05 & -   0.04) &   28.15 &(+    1.70 & -   1.73) &    0.08 &    2.69 & $\pm$   0.46 & upper limit     \cr
\noalign{\smallskip} \hline
\end{tabular}
\end{center}
\ \ Note. --
The meanings of columns are:  (1) central 
redshift of bin; (2) number of galaxies in bin;
(3) the value of the adopted faint-end
 slope; (4) the value of $M^*_B$, and upper (5) and 
lower (6) 68\% Poisson errors; (7) mean density
 $\phi^*$ followed by the 68\% Poisson errors 
 in columns (8) 
and (9); (10) square root of the fractional cosmic variance error, 
based on field geometry, bin volume and galaxy bias 
($b$) as a function of color (see text)
(11) luminosity density,  followed in (12) by a 
conservative error that combines 
Poisson errors in $M^*_B$ and $\phi^*$ 
with cosmic variance in quadrature;
 see text for further explanation.
(13) indicates whether the fits were 
calculated using the minimal, average or optimal weighting 
schemes, as described in \S3.3, or placing all failed 
red galaxies at $z$ = (0.7, 0.9, 1.1), as described in
 \S4.1
\end{table}

\begin{table}
\caption{Schechter function parameters for Blue galaxy samples}
\begin{center}
\begin{tabular}{crcrrrrrrrrcl}
\hline
\hline
$\langle z \rangle$ & N$_{gal}$ & $\alpha$ &\multicolumn{3}{c}{$M_B^*$} &\multicolumn{3}{c}{$\phi^*$} &$\sqrt{Var}$&\multicolumn{2}{c}{$j_B$} & Weights \cr
 {} & {} & {} & \multicolumn{3}{c}{ } &\multicolumn{3}{c}{$\times$ 10$^{-4}$ Gal Mpc$^{-3}$} & {} &\multicolumn{2}{c}{$\times$ 10$^{8}$ $L_{\odot}$}&{} \cr
 (1) & (2) & (3) & (4) & (5) & (6) & (7) & (8) & (9) & (10) & (11) & (12) & (13)\cr
\noalign{\smallskip} \hline \noalign{\smallskip}
      0.30 &   627 &  -1.30 &  -20.36 & (+   0.13 & -   0.11) &   31.78 &(+    2.15 & -   1.87) &    0.18 &    0.89 & $\pm$   0.20 & minimal         \cr
      0.50 &   812 &  -1.30 &  -20.72 & (+   0.05 & -   0.07) &   33.40 &(+    1.39 & -   1.77) &    0.16 &    1.31 & $\pm$   0.23 & minimal         \cr
      0.70 &   764 &  -1.30 &  -21.15 & (+   0.07 & -   0.07) &   24.67 &(+    1.35 & -   1.58) &    0.15 &    1.44 & $\pm$   0.26 & minimal         \cr
      0.90 &  2644 &  -1.30 &  -21.21 & (+   0.00 & -   0.03) &   27.27 &(+    0.35 & -   0.42) &    0.08 &    1.68 & $\pm$   0.13 & minimal         \cr
      1.10 &  1224 &  -1.30 &  -21.38 & (+   0.04 & -   0.05) &   20.84 &(+    1.08 & -   1.58) &    0.08 &    1.50 & $\pm$   0.16 & minimal         \cr
      1.30 &   448 &  -1.30 &  -21.86 & (+   0.07 & -   0.08) &   13.44 &(+    2.00 & -   2.71) &    0.07 &    1.51 & $\pm$   0.31 & minimal         \cr
{} &{} &{} &{} &{} &{} &{} &{} &{} &{} &{} &{} \cr
      0.30 &   627 &  -1.30 &  -20.19 & (+   0.10 & -   0.14) &   38.98 &(+    1.99 & -   2.73) &    0.18 &    0.94 & $\pm$   0.21 & average         \cr
      0.50 &   812 &  -1.30 &  -20.53 & (+   0.06 & -   0.09) &   44.07 &(+    2.05 & -   2.97) &    0.16 &    1.45 & $\pm$   0.27 & average         \cr
      0.70 &   764 &  -1.30 &  -21.04 & (+   0.05 & -   0.06) &   30.25 &(+    1.36 & -   1.25) &    0.15 &    1.59 & $\pm$   0.27 & average         \cr
      0.90 &  2644 &  -1.30 &  -21.14 & (+   0.03 & -   0.00) &   32.43 &(+    0.55 & -   0.35) &    0.08 &    1.87 & $\pm$   0.15 & average         \cr
      1.10 &  1224 &  -1.30 &  -21.33 & (+   0.03 & -   0.03) &   25.13 &(+    1.29 & -   1.01) &    0.08 &    1.72 & $\pm$   0.16 & average         \cr
      1.30 &   448 &  -1.30 &  -21.81 & (+   0.06 & -   0.06) &   16.39 &(+    2.55 & -   2.94) &    0.07 &    1.75 & $\pm$   0.33 & average         \cr
\noalign{\smallskip} \hline
\end{tabular}
\end{center}
\ \ Note. --
The meanings of columns are the same as in Table 3.
\end{table}

\begin{table}
\caption{Schechter function parameters for Red galaxy samples}
\begin{center}
\begin{tabular}{crcrrrrrrrrcl}
\hline
\hline
$\langle z \rangle$ & N$_{gal}$ & $\alpha$ &\multicolumn{3}{c}{$M_B^*$} &\multicolumn{3}{c}{$\phi^*$} &$\sqrt{Var}$&\multicolumn{2}{c}{$j_B$} & Weights \cr
 {} & {} & {} & \multicolumn{3}{c}{ } &\multicolumn{3}{c}{$\times$ 10$^{-4}$ Gal Mpc$^{-3}$} & {} &\multicolumn{2}{c}{$\times$ 10$^{8}$ $L_{\odot}$}&{} \cr
 (1) & (2) & (3) & (4) & (5) & (6) & (7) & (8) & (9) & (10) & (11) & (12) & (13)\cr
\noalign{\smallskip} \hline \noalign{\smallskip}
      0.30 &   109 &  -0.50 &  -21.02 & (+   0.18 & -   0.17) &   17.06 &(+    1.65 & -   1.64) &    0.26 &    0.60 & $\pm$   0.16 & minimal         \cr
      0.50 &   173 &  -0.50 &  -20.97 & (+   0.14 & -   0.10) &   14.15 &(+    0.70 & -   0.62) &    0.23 &    0.48 & $\pm$   0.10 & minimal         \cr
      0.70 &   196 &  -0.50 &  -21.19 & (+   0.06 & -   0.06) &   13.66 &(+    1.09 & -   1.00) &    0.22 &    0.56 & $\pm$   0.10 & minimal         \cr
      0.90 &   535 &  -0.50 &  -21.11 & (+   0.04 & -   0.05) &   10.72 &(+    0.38 & -   0.36) &    0.11 &    0.41 & $\pm$   0.04 & minimal         \cr
      1.10 &   178 &  -0.50 &  -21.44 & (+   0.07 & -   0.08) &    5.24 &(+    0.79 & -   0.95) &    0.11 &    0.27 & $\pm$   0.05 & minimal         \cr
{} &{} &{} &{} &{} &{} &{} &{} &{} &{} &{} &{} \cr
      0.30 &   109 &  -0.50 &  -20.86 & (+   0.16 & -   0.17) &   18.89 &(+    1.89 & -   1.85) &    0.26 &    0.58 & $\pm$   0.18 & average         \cr
      0.50 &   173 &  -0.50 &  -20.83 & (+   0.12 & -   0.09) &   17.71 &(+    1.03 & -   1.13) &    0.23 &    0.52 & $\pm$   0.13 & average         \cr
      0.70 &   196 &  -0.50 &  -21.05 & (+   0.06 & -   0.06) &   17.63 &(+    1.29 & -   1.50) &    0.22 &    0.64 & $\pm$   0.15 & average         \cr
      0.90 &   535 &  -0.50 &  -21.02 & (+   0.04 & -   0.02) &   13.47 &(+    0.60 & -   0.82) &    0.11 &    0.47 & $\pm$   0.06 & average         \cr
      1.10 &   178 &  -0.50 &  -21.33 & (+   0.08 & -   0.07) &    7.51 &(+    1.31 & -   1.52) &    0.11 &    0.35 & $\pm$   0.08 & average         \cr
{} &{} &{} &{} &{} &{} &{} &{} &{} &{} &{} &{} \cr
      0.70 &   334 &  -0.50 &  -20.79 & (+   0.08 & -   0.07) &   24.46 &(+    1.41 & -   1.71) &    0.22 &    0.70 & $\pm$   0.14 & upper limit     \cr
      0.90 &   848 &  -0.50 &  -20.89 & (+   0.03 & -   0.03) &   18.28 &(+    0.83 & -   0.38) &    0.11 &    0.57 & $\pm$   0.10 & upper limit     \cr
      1.10 &   548 &  -0.50 &  -21.43 & (+   0.05 & -   0.04) &   15.19 &(+    0.85 & -   1.21) &    0.11 &    0.78 & $\pm$   0.13 & upper limit     \cr
\noalign{\smallskip} \hline
\end{tabular}
\end{center}
\ \ Note. --
The meanings of columns are the same as in Table 3.
\end{table}

\begin{table}
\caption{Schechter function parameters for Moderately Blue galaxy sample}
\begin{center}
\begin{tabular}{crcrrrrrrrrcl}
\hline
\hline
$\langle z \rangle$ & N$_{gal}$ & $\alpha$ &\multicolumn{3}{c}{$M_B^*$} &\multicolumn{3}{c}{$\phi^*$} &$\sqrt{Var}$&\multicolumn{2}{c}{$j_B$} & Weights \cr
 {} & {} & {} & \multicolumn{3}{c}{ } &\multicolumn{3}{c}{$\times$ 10$^{-4}$ Gal Mpc$^{-3}$} & {} &\multicolumn{2}{c}{$\times$ 10$^{8}$ $L_{\odot}$}&{} \cr
 (1) & (2) & (3) & (4) & (5) & (6) & (7) & (8) & (9) & (10) & (11) & (12) & (13)\cr
\noalign{\smallskip} \hline \noalign{\smallskip}
      0.30 &   306 &  -1.30 &  -20.73 & (+   0.25 & -   0.22) &   13.42 &(+    1.18 & -   1.32) &    0.18 &    0.53 & $\pm$   0.16 & minimal         \cr
      0.50 &   440 &  -1.30 &  -20.96 & (+   0.10 & -   0.11) &   16.43 &(+    1.09 & -   0.78) &    0.16 &    0.80 & $\pm$   0.16 & minimal         \cr
      0.70 &   372 &  -1.30 &  -21.57 & (+   0.13 & -   0.11) &    9.58 &(+    0.59 & -   0.66) &    0.15 &    0.82 & $\pm$   0.16 & minimal         \cr
      0.90 &  1605 &  -1.30 &  -21.40 & (+   0.03 & -   0.03) &   14.59 &(+    0.29 & -   0.55) &    0.08 &    1.07 & $\pm$   0.09 & minimal         \cr
      1.10 &   846 &  -1.30 &  -21.49 & (+   0.05 & -   0.04) &   12.86 &(+    0.66 & -   0.52) &    0.08 &    1.02 & $\pm$   0.10 & minimal         \cr
      1.30 &   328 &  -1.30 &  -22.05 & (+   0.10 & -   0.08) &    7.42 &(+    1.51 & -   1.84) &    0.07 &    0.99 & $\pm$   0.25 & minimal         \cr
{} &{} &{} &{} &{} &{} &{} &{} &{} &{} &{} &{} \cr
      0.30 &   306 &  -1.30 &  -20.55 & (+   0.23 & -   0.22) &   16.48 &(+    1.60 & -   1.81) &    0.18 &    0.55 & $\pm$   0.16 & average         \cr
      0.50 &   440 &  -1.30 &  -20.73 & (+   0.11 & -   0.10) &   22.56 &(+    1.17 & -   1.24) &    0.16 &    0.89 & $\pm$   0.18 & average         \cr
      0.70 &   372 &  -1.30 &  -21.43 & (+   0.11 & -   0.13) &   12.08 &(+    0.81 & -   0.76) &    0.15 &    0.91 & $\pm$   0.18 & average         \cr
      0.90 &  1605 &  -1.30 &  -21.34 & (+   0.02 & -   0.01) &   16.97 &(+    0.26 & -   0.23) &    0.08 &    1.18 & $\pm$   0.09 & average         \cr
      1.10 &   846 &  -1.30 &  -21.43 & (+   0.04 & -   0.05) &   15.40 &(+    0.78 & -   1.09) &    0.08 &    1.16 & $\pm$   0.12 & average         \cr
      1.30 &   328 &  -1.30 &  -22.01 & (+   0.10 & -   0.11) &    8.70 &(+    1.83 & -   2.40) &    0.07 &    1.12 & $\pm$   0.30 & average         \cr
\noalign{\smallskip} \hline
\end{tabular}
\end{center}
\ \ Note. --
The meanings of columns are the same as in Table 3.
\end{table}

\begin{table}
\caption{Schechter function parameters for Very Blue galaxy sample}
\begin{center}
\begin{tabular}{crcrrrrrrrrcl}
\hline
\hline
$\langle z \rangle$ & N$_{gal}$ & $\alpha$ &\multicolumn{3}{c}{$M_B^*$} &\multicolumn{3}{c}{$\phi^*$} &$\sqrt{Var}$&\multicolumn{2}{c}{$j_B$} & Weights \cr
 {} & {} & {} & \multicolumn{3}{c}{ } &\multicolumn{3}{c}{$\times$ 10$^{-4}$ Gal Mpc$^{-3}$} & {} &\multicolumn{2}{c}{$\times$ 10$^{8}$ $L_{\odot}$}&{} \cr
 (1) & (2) & (3) & (4) & (5) & (6) & (7) & (8) & (9) & (10) & (11) & (12) & (13)\cr
\noalign{\smallskip} \hline \noalign{\smallskip}
      0.30 &   321 &  -1.30 &  -19.96 & (+   0.11 & -   0.12) &   19.84 &(+    1.95 & -   2.04) &    0.18 &    0.39 & $\pm$   0.09 & minimal         \cr
      0.50 &   372 &  -1.30 &  -20.38 & (+   0.12 & -   0.11) &   18.84 &(+    1.91 & -   1.71) &    0.16 &    0.54 & $\pm$   0.12 & minimal         \cr
      0.70 &   403 &  -1.30 &  -20.63 & (+   0.12 & -   0.12) &   18.45 &(+    1.58 & -   1.71) &    0.15 &    0.67 & $\pm$   0.14 & minimal         \cr
      0.90 &  1211 &  -1.30 &  -20.84 & (+   0.06 & -   0.03) &   15.06 &(+    0.82 & -   0.65) &    0.08 &    0.66 & $\pm$   0.07 & minimal         \cr
      1.10 &   621 &  -1.30 &  -20.94 & (+   0.08 & -   0.07) &   12.35 &(+    0.82 & -   1.03) &    0.08 &    0.59 & $\pm$   0.07 & minimal         \cr
      1.30 &   670 &  -1.30 &  -21.02 & (+   0.04 & -   0.05) &   21.87 &(+    1.26 & -   1.19) &    0.07 &    1.13 & $\pm$   0.12 & minimal         \cr
{} &{} &{} &{} &{} &{} &{} &{} &{} &{} &{} &{} \cr
      0.30 &   321 &  -1.30 &  -19.79 & (+   0.12 & -   0.15) &   24.39 &(+    2.28 & -   2.25) &    0.18 &    0.41 & $\pm$   0.10 & average         \cr
      0.50 &   372 &  -1.30 &  -20.22 & (+   0.14 & -   0.13) &   23.63 &(+    2.18 & -   3.03) &    0.16 &    0.59 & $\pm$   0.14 & average         \cr
      0.70 &   403 &  -1.30 &  -20.51 & (+   0.13 & -   0.11) &   22.57 &(+    2.36 & -   2.08) &    0.15 &    0.73 & $\pm$   0.15 & average         \cr
      0.90 &  1211 &  -1.30 &  -20.72 & (+   0.04 & -   0.05) &   19.49 &(+    1.11 & -   1.38) &    0.08 &    0.76 & $\pm$   0.08 & average         \cr
      1.10 &   621 &  -1.30 &  -20.84 & (+   0.07 & -   0.08) &   16.89 &(+    1.33 & -   1.60) &    0.08 &    0.74 & $\pm$   0.10 & average         \cr
      1.30 &   670 &  -1.30 &  -20.86 & (+   0.05 & -   0.04) &   38.92 &(+    2.06 & -   2.16) &    0.07 &    1.73 & $\pm$   0.17 & average         \cr
\noalign{\smallskip} \hline
\end{tabular}
\end{center}
\ \ Note. --
The meanings of columns are the same as in Table 3.
\end{table}

\clearpage

\begin{appendix}
\section{K-Corrections}

The luminosity functions in this paper use Johnson rest-frame $B$ and
$U-B$ magnitudes and colors. Since $B_{Johnson}$ matches
observed $B$, $R$ and $I$ only at certain redshifts, the
transformation into rest-frame quantities requires the calculation of 
K-corrections (e.g., Oke \& Sandage 1962; Hogg et al. 2002). Because of
the rather limited number of bands (3 for DEEP2), the
use of more robust techniques for the calculation of K-corrections 
as employed by COMBO-17 (Wolf et al. 2003) or SDSS (Blanton et al. 2003)
is not possible. The procedure in this work is
similar to that of Gebhardt et al. (2003), 
who used nearby galaxy SEDs from Kinney et
al. (1996) to relate the observed color and magnitude at redshift $z$
to the rest-frame color and $B$-band magnitude.

We started with the 43 Kinney et al. SEDs 
whose spectra cover the range 1,100 \AA~$\leq \lambda \leq$
10,000 \AA~ without gaps, as listed in
Table A8. Even though the Kinney et al. 
spectra are integrated only over a small aperture (10$'' \times
20''$) (in contrast to DEEP2 galaxy magnitudes and colors, which are
close to total), this approach was chosen in preference to model
spectra because the Kinney et al. data represent
{\it real} spectra. 
The convolution between filter responses and galaxy SEDs
followed Fukugita, Shimasaku \& Ichikawa (1995) by resampling
filters and spectra to the same dispersion (1 \AA), 
using parabolic and 
linear interpolations respectively.  
Still following Fukugita et al. (1995), the  curves for Johnson $U$ and $B$
filters come from Buser (1978) and  Azusienis \& Straizys (1969)
respectively.
The
throughput curves for the 
CFHT $12k\times8k$ DEEP2 $BRI$ imaging were calculated by Nick Kaiser, who
provided filter transmission curves, 
CCD quantum efficiency curves, and the telescope response function. Normalized
curves for the CFHT filters are shown in Figure A11. Calibration of
these convolutions used the model atmosphere of Vega calculated by
Kurucz that is distributed with the Bruzual \& Charlot (2003) galaxy evolution
synthesis package. The conversion between Vega and AB magnitudes
(Table 1) simply compared the zero-points between the Vega calibration and
that obtained using a flat spectrum in F($\nu$) converted into
wavelength space (e.g., Fukugita et al. 1995).

Figure A12 compares synthesized $U-B$ values for 
the Kinney et al. galaxies with $U-B$ values for the same galaxies 
derived from
the {\it Third Revised Catalog of Galaxies} (de Vaucouleurs et
al. 1991, RC3). The latter were calculated using the RC3 raw total $U-B$
colors, corrected only for Galactic absorption using the Schlegel et
al. (1998) extinction values tabulated in the NASA Extragalactic
Database\footnote{{\it http://nedwww.ipac.caltech.edu}}. Both
sets of measurements  are therefore consistent in being corrected for Galactic
extinction though not for internal absorption or for a face-on
geometry.
The agreement is fairly good, even
though the RC3 values refer to $total$ galaxy colors
while the Kinney et al. spectra sample the center only. 
For the reddest templates, the synthetic spectra overerestimate $U-B$
by $\sim$0.08 mag; this difference is in the expected direction of the
natural internal color gradient. 
Overall, the good agreement in Figure A12 suggests that the zero-point
of our synthetic $U-B$ system is accurate
to a few hundredths of a magnitude.

Figure A13 shows the calculated K-correction $K_{RB}$ (which converts 
$R$ into $B_{Johnson}$)  as a function of synthetic observed 
$R-I$ color for
different redshift intervals, while Figure A14 shows calculated rest-frame 
$U-B$ as a function of 
synthetic observed $B-R$ in the same redshift intervals.
Similar curves of $U-B$ as a function of observed $R-I$
for DEEP2 galaxies, are shown in Figure A15.

In general, relations are tight at redshifts where $U$ and $B$
are shifted close to the observed passbands but show
more scatter as the match worsens.
For redshifts beyond $\sim$0.7,
where DEEP2 is focused, $R-I$ color provides a much better
estimate of rest-frame $U-B$ and $B$ than $B-R$.

Finally, Figure A16 compares synthetic DEEP2 $B-R$ versus $R-I$
colors from the Kinney et al. (1996) SEDs versus real data, binned by
redshift.   Observed galaxies are  
the red and green data points, while synthetic colors from the 
Kinney et al. templates are the black triangles; only
34 templates (identified in Table A1)  are displayed here.
A similar diagram
using the whole set of 43 templates was used to select the final set. 
Whenever a template 
was an outlier compared to the observed galaxy distribution, it was flagged;
templates flagged in more than two redshift bins were discarded. 
Galaxies that were discarded have a ``no'' in column (4) in Table A1 and are
shown as asterisks in Figures A2 through A5.

The good agreement between observed and synthesized colors in
Figure A16
suggests that, even though evolution of the template SEDs is being
neglected in the present K-corrections, the errors introduced are
probably small. 
A reason for this is that the observed color range of galaxies at all
redshifts considered in this work is well covered by the spectral locus
of the templates.
A possible shortcoming of not using evolving SEDs for the
K-correction, i.e., K+e corrections, 
is that at higher redshifts a portion of galaxies
might shift into the wrong color class, as discussed by Wolf et al. (2003)
and Bell et al. (2004).  This problem is avoided in the present work by
dividing galaxies into red and blue classes using the evolving
``valley'' of color bimodality. This does not prevent galaxies from
changing color---indeed,  
the number of red galaxies may
grow as blue galaxies migrate across the valley after
star-formation quenching---but it does define classes of
galaxies in a way that is independent of color zero-point errors.

Second-order polynomials were used to estimate $U-B$ and the
K-corrections from the observed colors.  Custom fits were
calculated (at the specific redshift of each observed galaxy)
of $U-B$ and the K-correction versus   $B-R$  and/or $R-I$.
Rest-frame parameters were
obtained by entering the observed colors.
The range of estimated colors (and
K-corrections) was restricted to that covered by the template
spectra, so that observed
galaxies with extreme colors 
were forced to have reasonable 
rest-frame values. For DEEP2 galaxies, at redshifts where rest-frame
$U-B$ lies between $B-R$ and $R-I$, the K-corrections and
rest-frame colors were obtained by interpolating between the $B-R$ and
$R-I$ derived quantities. Otherwise, the rest-frame quantities were
obtained using the closest pair of filters.
The RMS error for estimated $U-B$ ranges from
0.12 mag at $z = 1.2$ (worst value) to  0.03 mag at 
redshifts where the observed filters best overlap $U-B$. 
The RMS error in $K_{RB}$ ranges from $\sim$0.01 mag
whenever one of the observed filters overlaps $B_{Johnson}$ to  
$\sim$0.15 mag at $z\sim 1.5$, where a large extrapolation is
being used.
The results obtained using the parabolic fits are comparable
to the results using interpolations between SEDs (Lilly et
al. 1995).

This procedure differs from that of Gebhardt et al. (2003) in two 
ways.  First, Gebhardt et al. used nearly all the
Kinney et al. (1996) templates after removing only two very deviant
spectra. Second, the parabolic fit here 
between observed and rest-frame parameters is calculated at the
exact redshift of each observed galaxy, whereas Gebhardt et
al. attempted to calculate a more
general polynomial that mapped the 
color transformation over the entire redshift range.
\begin{figure}\label{A11}
\vspace{60mm}
\includegraphics{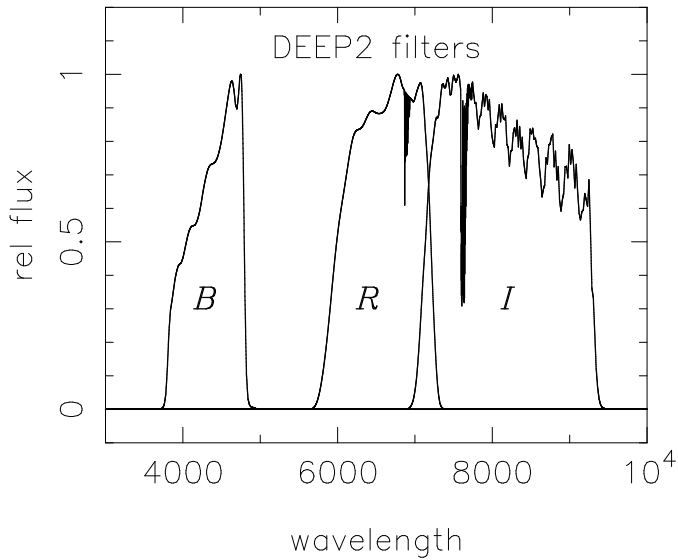}
\caption{Transmission curves of the CFHT
  $12K\times8K$ filter transmission curves (including telescope and
  CCD throughput) used in DEEP2. Also shown are estimates of the
  telluric absorption due to the A ($\sim$ 6800 \AA) and B bands
  ($\sim$ 7600 \AA).
}
\end{figure}
%
%
\begin{figure}\label{A12}
\vspace{80mm}
\includegraphics{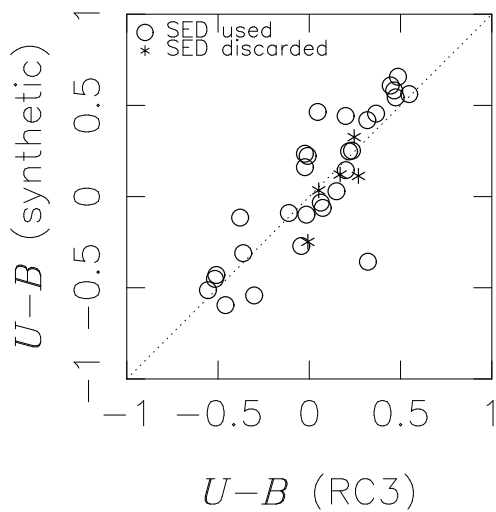}
\caption{Comparison between $U-B$ colors derived from
  the RC3, with synthesized $U-B$ obtained by convolving
  Johnson filters with the Kinney et al. (1996) spectra.
  The synthetic $U-B$ are based on the template SEDs, which
   are corrected for Galactic absorption only. To match these, the
  RC3 $U-B$ colors have been corrected for
  Galactic absorption only (using Schlegel et al. 1998) but not
  for internal absorption.
  Galaxies used in the final K-correction fits are shown as  
  circles, while galaxies whose spectra were discarded from the final
  fits are shown as asterisks. The RC3 measurements are total (containing
  all the galaxy light), while the Kinney et
  al. spectra sample only a rectangular 20$''\times10''$ box at the 
  center of the galaxy. In spite of this, the deviations from the
  dotted line are fairly small.
}
\end{figure}
\begin{figure}\label{A13}
\vspace{90mm}
\includegraphics{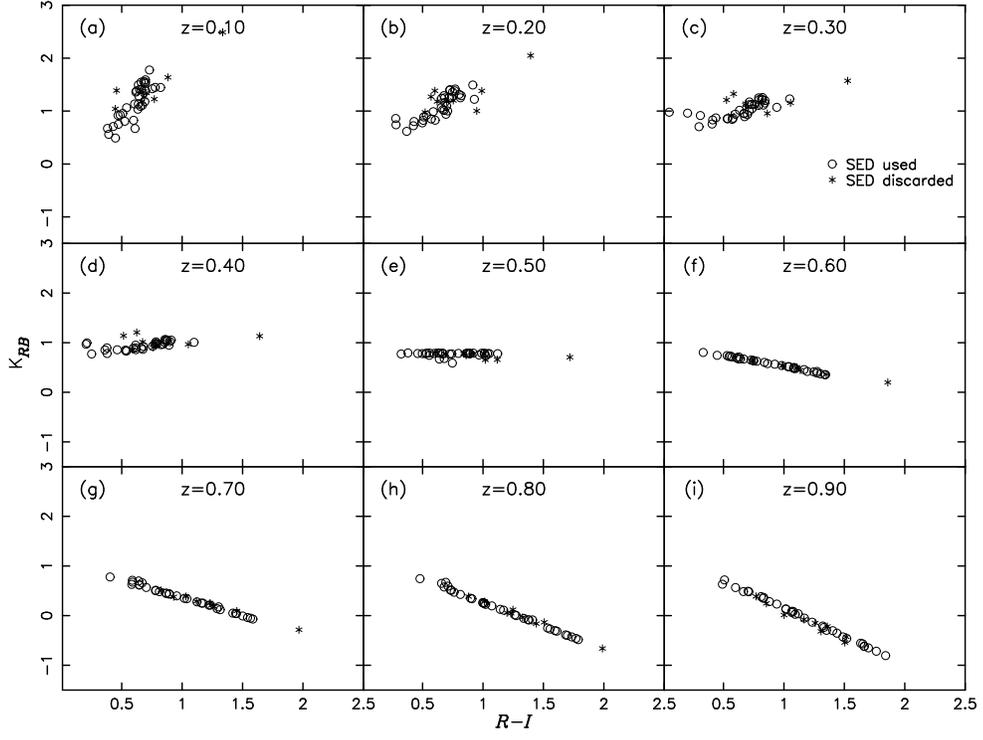}
\caption{The K-correction transforming DEEP2
  $R$ to $B_{Johnson}$ as a function
  of SED color and redshift $z$ for all Kinney et al.
  template galaxies in Table A1. 
  SED color is apparent DEEP2 $R-I$ synthesized from the Kinney et
  al. (1996) templates at that redshift.
  Galaxies used in the final fit are shown as open
  circles, while galaxies removed from the final fit are shown as
  asterisks. At $z\sim$ 0.5, $R$ and $B_{Johnson}$ essentially
  overlap, and there is little dependence of $K_{RB}$ on observed color.
}
\end{figure}
\begin{figure}\label{A14}
\vspace{80mm}
\includegraphics{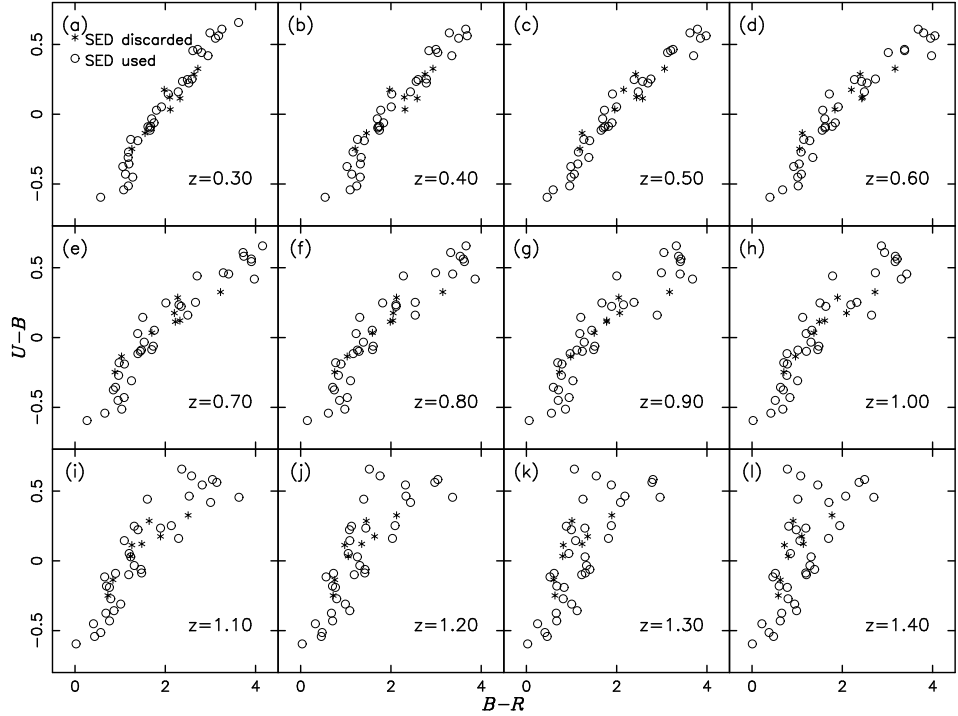}
\caption{Similar to Figure A3, but showing synthesized $U-B$ color as
  a function of synthesized observed $B-R$, versus redshift for the
  Kinney et al. SEDs. As the
  overlap between redshifted 
  $U-B$ color and observed color decreases towards higher redshifts,
  the relation becomes noisier.
}
\end{figure}
%
%
\begin{figure}\label{A15}
\vspace{90mm}
\includegraphics{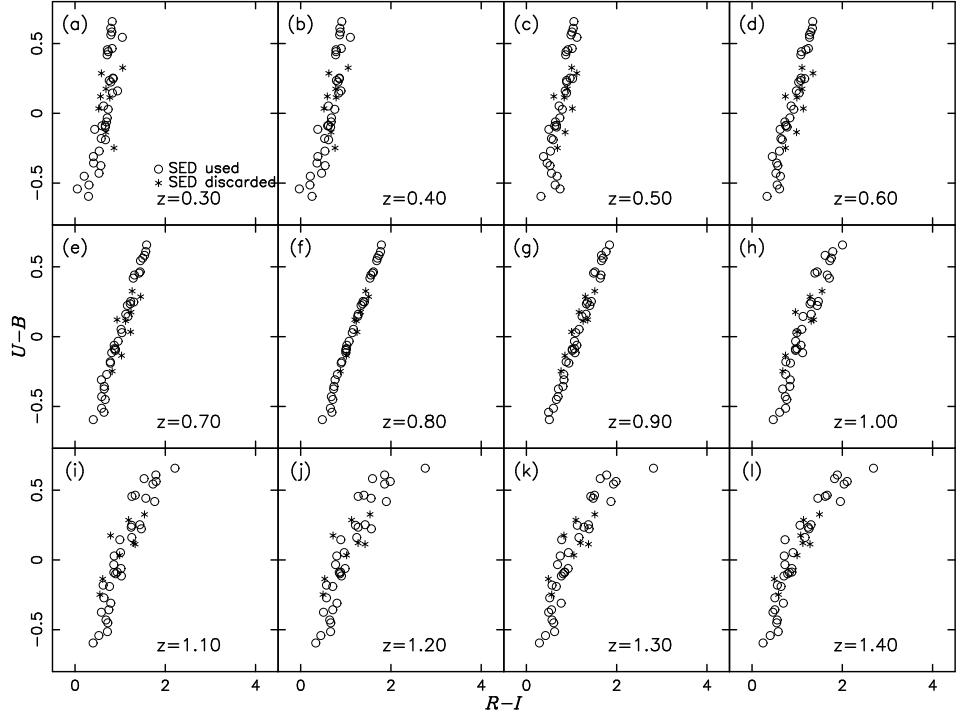}
\caption{Similar to Figure A4, but showing synthesized rest-frame 
 $U-B$ color as a function of synthesized observed
  $R-I$, used in DEEP2, versus redshift. In contrast to the $B-R$
  plots shown in Figure A4, the relation between $R-I$ and $U-B$
  gets tighter at high redshift, where $R-I$ is a better estimator of
  rest-frame color. 
}
\end{figure}
%
%
\begin{figure}\label{A16}
\vspace{80mm}
\includegraphics{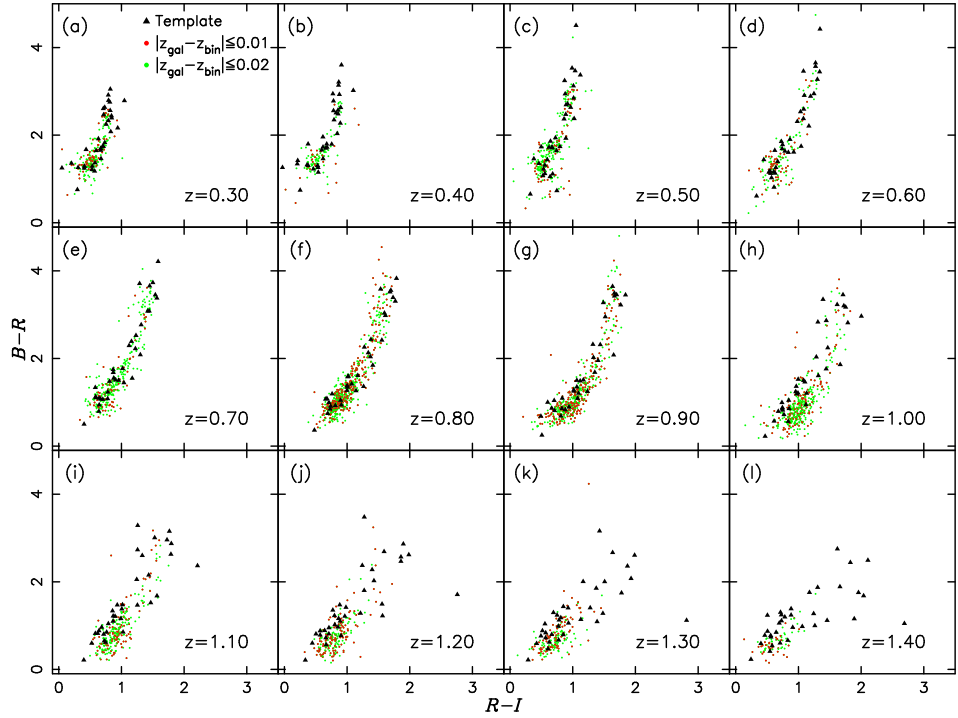}
\caption{Comparison between synthesized $R-I$ and $B-R$ colors
  measured from the final sample of 34 Kinney et al.
  templates (black triangles) versus observed colors of DEEP2 galaxies.
  Red dots represent DEEP2 galaxies with good quality redshifts lying
  within $\pm$0.01 of the redshift displayed in the plot, while green
  dots represent galaxies lying within $\pm$0.02 of 
  that redshift. Even though the template SEDs are not
  evolved, they still provide a good match to the observed data, even at the
  higher ranges of the sample.
}
\end{figure}
\begin{table}
\smallskip
\caption{Kinney et al. SEDs}
\begin{center}
\begin{tabular}{lrrc}
\hline
\hline
Id &$U-B$ (RC3)  &$U-B$ (synthetic) & Used in Analysis\cr
\noalign{\smallskip} \hline \noalign{\smallskip}
NGC~5128        & \nodata &  0.80  &no\cr
NGC~1399        &  0.49   &  0.66  &yes\cr
NGC~7196        &  0.45   &  0.61  &yes\cr
NGC~1553        &  0.47   &  0.58  &yes\cr
NGC~1404        &  0.55   &  0.56  &yes\cr
NGC~4594        &  0.47   &  0.54  &yes\cr
NGC~210         &  0.05   &  0.46  &yes\cr
NGC~1433        &  0.20   &  0.44  &yes\cr
NGC~1316        &  0.37   &  0.46  &yes\cr
NGC~2865        &  0.32   &  0.42  &yes\cr
NGC~1808        &  0.25   &  0.33  &no\cr
NGC~3393        & \nodata &  0.29  &no\cr
NGC~7582        &  0.23   &  0.25  &yes\cr
NGC~3081        &  0.22   &  0.25  &yes\cr
NGC~7083        & -0.02   &  0.24  &yes\cr
NGC~7590        & -0.01   &  0.22  &yes\cr
NGC~3660        & \nodata &  0.17  &no\cr
NGC~6221        & -0.02   &  0.16  &yes\cr
NGC~1097        &  0.20   &  0.15  &yes\cr
NGC~5102        &  0.17   &  0.12  &no\cr
NGC~1326        &  0.27   &  0.11  &no\cr
IC~3639         & \nodata &  0.05  &yes\cr
NGC~1068        &  0.05   &  0.03  &no\cr
NGC~3351        &  0.15   &  0.03  &yes\cr
NGC~5135        &  0.06   & -0.03 &yes\cr
NGC~7552        &  0.07   & -0.06 &yes\cr
NGC~7130        & \nodata & -0.09 &yes\cr
NGC~7793        & -0.11   & -0.09 &yes\cr
NGC~1672        & -0.02   & -0.10 &yes\cr
NGC~7673        & -0.38   & -0.12 &yes\cr
NGC~4748        & \nodata & -0.14 &no\cr
CGCG~038-052    & \nodata & -0.18 &yes\cr
NGC~7496        & \nodata & -0.19 &yes\cr
NGC~4385        & -0.01   & -0.25 &no\cr
M~83            & -0.04   & -0.27 &yes\cr
NGC~1313        & -0.36   & -0.31 &yes\cr
ESO~296~G~11    &  0.32   & -0.36 &yes\cr
NGC~3049        & \nodata & -0.38 &yes\cr
NGC~7714        & -0.51   & -0.43 &yes\cr
Tololo 1924-416 & -0.52   & -0.45 &yes\cr
NGC~3125        & -0.56   & -0.51 &yes\cr
NGC~5253        & -0.30   & -0.54 &yes\cr
NGC~1705        & -0.46   & -0.60 &yes\cr
\noalign{\smallskip} \hline
\end{tabular}
\end{center}
\ \ Note. --
SEDs of galaxies that are used in the calculation of 
K-corrections are denoted by ``yes'' in column 4. Galaxies that
were discarded because of deviant behavior in two or more redshift
intervals are noted by ``no.'' These are represented as
asterisks in Figure A2 (if RC3 data exist) and in Figures A3 through A6.
\end{table}

\clearpage
\section{Stability of Schechter Parameters as a Function of Faint
  Limiting Magnitude} 

A limitation that is invariably present when calculating the
luminosity function of galaxies is the smaller domain accessible
in absolute
magnitudes at higher redshifts. To examine 
this effect, Schechter fits were re-calculated for 
DEEP2 red galaxies considering three different lower-$z$
bins and raising the faint limit to brighter magnitudes to
match the magnitude ranges accessible in the higher-$z$ bins.
Any mismatch in the assumed shape of the luminosity function will
result in a spurious drift of the fitted parameters
as the magnitude limit is raised.  The purpose of this test is to
make sure that our measured evolutions in $M^*_B$ and $\phi^*$ for red
galaxies are not contaminated by this kind of bias.

The results of this test are shown in Figure B17 for red galaxies, where $\alpha$
has been kept at the value $-0.5$ used in the main text.  Vertical
gray lines show the limits of the data in bins $z = $0.8-1.0
and $z = $1.0-1.2.   For all three lower bins, we see a   
drift of $M^*_B$ of $\sim$0.1 mag toward {\it fainter} values as
the samples are truncated, whereas the measured evolution is
a {\it brightening} of $M_B^*$ back in time.  Thus, if anything,
the true evolution in $M_B^*$ is slightly more than 
claimed.  The quantity $\phi^*$ drifts upward by 0.1-0.15 dex
with more truncation whereas the observed effect is a fall back in
time, so again, the true evolution may be underestimated.
Finally, the important quantity $j_B$ drifts
upward by only 0.05-0.1 dex, confirming its essentially
constant nature. A similar study was made for the blue galaxy sample,
which shows the same behavior. This test also shows that 1/$V_{max}$
seems to provide a more robust estimate of the Schechter parameters
than STY as the domain in absolute magnitudes decreases, though both
agree within the estimated errors.

We conclude that errors caused by Schechter function mismatches
are in all cases small compared to the measured evolutionary
changes for red galaxies.

\begin{figure}
\vspace{80mm}
\includegraphics{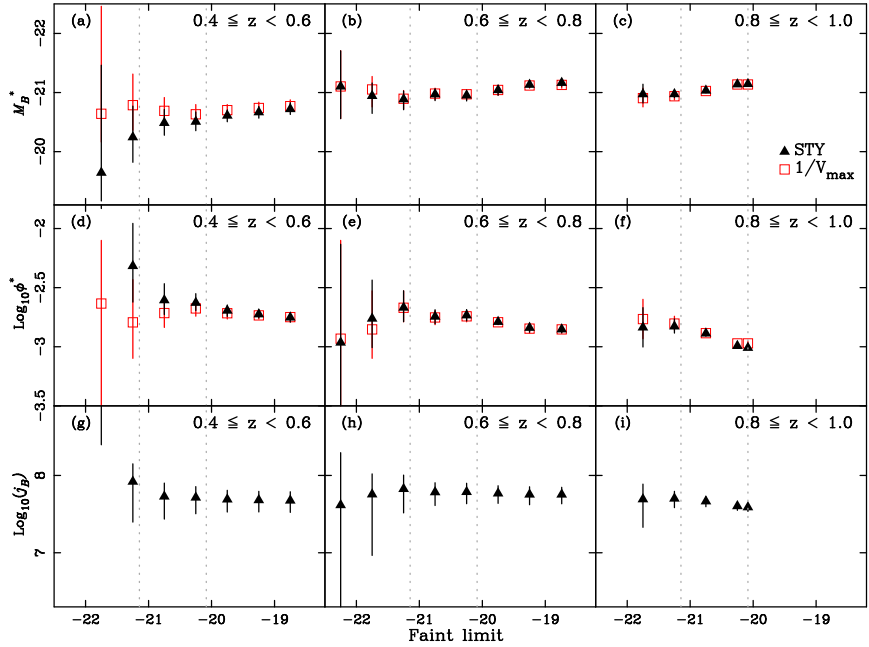}
\caption{
Stability of  Schechter parameters for red galaxies from DEEP2
as a function of the adopted faint
magnitude limit used in the fit. This is calculated in three different 
redshift ranges, where for each we have limited 
the data at progressively brighter
magnitudes. The vertical grey lines present the actual limits of the
data in bins $z = $0.8-1.0 (right line) and $z = $1.0-1.2 (left line).
Solid black triangles represent fits using the STY
method, and open red squares the 1/$V_{max}$ fits. Panels $a$-$c$
show results for $M^*_B$,  $d$-$f$ 
results for $\phi^*$, 
and $g$-$i$ results for the integrated luminosity density.
All fits use
constant $\alpha=-0.5$, as in the text. 
In the lowest-redshift bin, there are very few
galaxies brighter than $-21.5$, and fits that are highly truncated are poor.
Aside from this, the fits are quite stable, indicating that
drifts induced by a mismatch in the shape of the Schechter
function to the data are small. If anything, the trends here would
only $add$ to the observed trends. Appendix B provides a more
quantitative discussion. 
}
\end{figure}
\end{appendix}
\end{document}